%% file: paper.tex
\documentclass[useAMS,usenatbib]{mn2e}

\usepackage{aas_macros}
\usepackage{graphicx}
\usepackage{amsmath}
\usepackage{xspace}
\usepackage{ifthen}
\usepackage{lscape}
\usepackage{placeins}

\newcommand{\kms}{\mbox{km\,s$^{-1}$}\xspace}
\newcommand{\teff}{$T_\text{eff}$\xspace}    
\newcommand{\logg}{$\log g$\xspace}
\newcommand{\feh}{[Fe/H]\xspace}

\newcommand{\chisq}{$\chi ^2$\xspace}
\newcommand{\vsini}{$v \sin i_\text{rot}$\xspace}
\newcommand{\iraf}{\footnote{IRAF is distributed by the National Optical Astronomy Observatories, which are operated by the Association of Universities for Research in Astronomy, Inc., under cooperative agreement with the National Science Foundation.}}

\title[Transiting very low mass stars from HATSouth]{The Mass-Radius Relationship for Very Low Mass Stars:  Four New Discoveries from the HATSouth Survey
\thanks{The HATSouth network is operated by a collaboration consisting of
Princeton University (PU), the Max Planck Institute f\"ur Astronomie
(MPIA), and the Australian National University (ANU).  The station at
Las Campanas Observatory (LCO) of the Carnegie Institute, is operated
by PU in conjunction with collaborators at the Pontificia Universidad
Cat\'olica de Chile (PUC), the station at the High Energy Spectroscopic
Survey (HESS) site is operated in conjunction with MPIA, and the
station at Siding Spring Observatory (SSO) is operated jointly with
ANU.} }

\author[G.~Zhou et al.]
{\parbox{\textwidth}
{G.~Zhou$^{1}$\thanks{E-mail: \texttt{george.zhou@anu.edu.au}},
 D.~Bayliss$^{1}$,  
J.~D.~Hartman$^{2,3}$, 
G.~\'A.~Bakos$^{2,3}$\thanks{Alfred P.~Sloan Research Fellow}\thanks{Packard Fellow}, 
K.~Penev$^{2,3}$, 
Z.~Csubry$^{2,3}$, 
T.G.~Tan$^{4}$,  
A.~Jord\'an$^{5}$,  
L.~Mancini$^{6}$, 
M.~Rabus$^{5}$, 
R.~Brahm$^{5}$, 
N.~Espinoza$^{5}$, 
M.~Mohler-Fischer$^{6}$, 
S.~Ciceri$^{6}$, 
V.~Suc$^{5}$, 
B.~Cs\'ak$^{6}$, 
T.~Henning$^{6}$, and 
B.~Schmidt$^{1}$\vspace{0.4cm}}\\
\parbox{\textwidth}{
$^{1}${Research School of Astronomy and Astrophysics, Australian National University, Canberra, ACT 2611, Australia}\\
$^{2}${Department of Astrophysical Sciences,
	Princeton University, NJ 08544, USA}\\
$^{3}${Harvard-Smithsonian Center for Astrophysics,
	Cambridge, MA, USA}\\
$^{4}${Perth Exoplanet Survey Telescope, Perth, Australia}\\
$^{5}${Instituto de Astrof\'isica, Pontificia Universidad
 Cat\'olica de Chile, Av.\ Vicu\~na Mackenna
   4860, 7820436 Macul, Santiago, Chile}\\
$^{6}${Max Planck Institute for Astronomy, K\"{o}nigstuhl 17,
69117 -- Heidelberg, Germany}}}

\begin{document}

\date{Accepted 2013 October 27.  Received 2013 October 24; in original form 2013 October 4}

\pagerange{\pageref{firstpage}--\pageref{lastpage}} \pubyear{2002}

\maketitle

\label{firstpage}

\begin{abstract}
We report the discovery of four transiting F-M binary systems with companions between $0.1-0.2\,M_\odot$ in mass by the HATSouth survey. These systems have been characterised via a global analysis of the HATSouth discovery data, combined with high-resolution radial velocities and accurate transit photometry observations. We determined the masses and radii of the component stars using a combination of two methods: isochrone fitting of spectroscopic primary star parameters, and equating spectroscopic primary star rotation velocity with spin-orbit synchronisation. These new very low mass companions are HATS550-016B ($0.110_{-0.006}^{+0.005}\,M_\odot$, $0.147_{-0.004}^{+0.003}\,R_\odot$), HATS551-019B ($0.17_{-0.01}^{+0.01}\,M_\odot$, $0.18_{-0.01}^{+0.01}\,R_\odot$), HATS551-021B ($0.132_{-0.005}^{+0.014}\,M_\odot$, $0.154_{-0.008}^{+0.006}\,R_\odot$), HATS553-001B ($0.20_{-0.02}^{+0.01}\,M_\odot$, $0.22_{-0.01}^{+0.01}\,R_\odot$). We examine our sample in the context of the radius anomaly for fully-convective low mass stars. Combining our sample with the 13 other well-studied very low mass stars, we find a tentative 5\% systematic deviation between the measured radii and theoretical isochrone models.
\end{abstract}


\section{Introduction}
\label{sec:introduction}

`Very low mass stars' (VLMS), with masses between 0.08 and $0.3\,M_\odot$, are the most dominant subset of the stellar population \citep[e.g.][]{2001MNRAS.322..231K}. These stars are thought to have fully convective interiors and hydrogen fusion in their cores, distinguishing them from higher mass stars and brown dwarfs, respectively \citep[see review by][]{2000ARA&amp;A..38..337C}. Mass and radius are two of the most fundamental measurements for stars. Previous studies have shown that the radii of sub-solar mass stars are under-predicted by theoretical interior models at the $5-10$\% level \citep[e.g.][]{2002ApJ...567.1140T,2006Ap&amp;SS.304...89R,2010A&amp;ARv..18...67T,2012ApJ...757...42F,2013ApJ...776...87S}. The interior structure of the fully convective VLMSs is different to that of higher mass, partially radiative stars, and therefore warrants a more thorough, independent examination.

The vast majority of masses and radii come from dynamical measurements of binary systems. One explanation for the radius anomaly is that these M-dwarf binaries are spun-up by tidal interactions, the speed-up of the internal dynamo then leads to increased magnetic activities, suppressing convection and increasing star-spot activity \citep[e.g.][]{2006Ap&amp;SS.304...89R,2005ApJ...631.1120L,2007ApJ...660..732L,2010ApJ...718..502M}. \citet{2007A&amp;A...472L..17C} showed that the general radius discrepancies for low mass stars can be accounted for by allowing large spot coverages and varied mixing length in the models. However, since the magnetic field is thought to be generated differently in fully convective stars \citep[e.g.][]{2006A&amp;A...446.1027C}, the effect of this spin-up on the VLMSs and the resulting spot coverage is unclear. In addition, the effect of the mixing length parameter incorporated in stellar models for fully convective, near adiabatic, stars is significantly less than that for higher mass stars. Any explanation for the radius anomaly should also not be restricted to binaries, since the radius inflation is also observed for isolated M-dwarfs measured via interferometry \citep[e.g.][]{2006ApJ...644..475B,2012ApJ...757..112B,2013ApJ...776...87S}.

It remains difficult to test stellar models for the VLMS population, given that metallicities and precise (better than 10\%) mass radius measurements are available for only 13 previous objects (see Section~\ref{sec:mass-radi-relat}). In contrast, we know the masses and radii of $\sim 40$ exoplanets to better than 5\% precision, which has led to more thorough examinations of planet interior models \citep[e.g.][]{2011ApJ...729L...7L,2012ApJ...744...59S}. The radii of stand-alone, close-by M-dwarfs can be measured via interferometry \citep[e.g.][]{2003Aamp;A...397L...5S}, but the masses must be inferred from empirical mass-luminosity relationships. Dynamical masses of binaries can be obtained via astrometric orbit measurements \citep[e.g.][]{2013ApJ...773...28S}. Double-lined M-M eclipsing binary systems provide accurate, model-independent mass and radius measurements \citep{1996ApJ...456..356M,2011Sci...331..562C,2011Sci...333.1602D,2011ApJ...742..123I,2013MNRAS.tmp.1072N}, but these systems are relatively rare. In addition, the accuracy of M-M binary derived system parameters may suffer from M-dwarf activity and unaccounted spot variability, and may not be as reliable as previously thought \citep{2012ApJ...757...42F}.

Photometric transit surveys have lead to a rapid expansion in the population of transiting exoplanets. VLMSs have radii comparable to that of gas-giant planets, and are often found as companions in binary systems to solar-type stars. These F-M binaries exhibit similar transit signals as hot-Jupiter systems, and can be easily identified by transiting planet surveys. The population of well characterised VLMSs can be greatly extended by including single lined F-M binaries \citep[e.g.][]{2005A&amp;A...433L..21P,2006A&amp;A...447.1035P,2007ApJ...663..573B,2009ApJ...701..764F,2013Aamp;A...549A..18T}. 

There are a number of approaches towards measuring the mass and radius of M-dwarf companions in single lined F-M binary systems. The primary star properties can be obtained by combining spectroscopic analysis with stellar evolution models. The precision of the measured companion mass and radius are limited by the uncertainty in the primary star properties. For orbital companions of substantial mass, the rotation of the primary star is quickly synchronised with the companion orbital period. Fundamental system parameters derived from transit light curves, combined with rotational velocities measured from spectra, can yield relatively model-independent masses and radii for both components of a binary system \citep[e.g.][]{2007ApJ...663..573B,2009ApJ...701..764F}.

In this study, we present the discovery of four single-lined stellar systems with 0.1--0.2\,$M_\odot$ VLMS companions. These low mass eclipsing binaries were identified by the HATSouth survey \citep{2013PASP..125..154B}. The discovery and follow-up observations are detailed in Section~\ref{sec:observations}. Analysis of the individual systems, including spectral classifications of the primary star, global modelling of the light curves and radial velocity data, and descriptions of the methods used to derive the mass and radius of the companions, can be found in Section~\ref{sec:analysis}. Section~\ref{sec:discussion} discusses these new discoveries in the context of existing VLMS systems, and examines the mass-radius anomaly in the VLMS regime. 

\section{Observations}
\label{sec:observations}

\subsection{HATSouth photometric detection}
\label{sec:hatsouth-detection}

The transiting VLMS systems were identified from photometric observations by the HATSouth global network. HATSouth consists of six telescope units spread over three sites, Siding Spring Observatory (SSO) in Australia, Las Campanas Observatory (LCO) in Chile, and the HESS site in Namibia, providing continuous monitoring of 128 deg$^2$ fields in the Southern sky \citep{2013PASP..125..154B}. Each unit consists of four 0.18\,m f/2.8 Takahasi astrographs and  Apogee Alta-U16M D9 $4\text{k}\times4\text{k}$ front illuminated CCD cameras, with $9\,\mu\text{m}$ pixels, and plate scale of $3.7" \, \text{pixel}^{-1}$. The four telescopes are offset by $4^\circ$, allowing four adjacent $4^\circ \times 4^\circ$ fields to be simultaneously monitored. The observations are made at 4 minute cadence in the $r'$ band. Each field is monitored for $\sim 2$ months by a unit at each HATSouth station. Aperture photometry is performed on the reduced frames, and detrended using External Parameter Decorrelation \citep[EPD,][]{2007ApJ...670..826B} and Trend Filtering Algorithm \citep[TFA,][]{2005MNRAS.356..557K}. Objects exhibiting periodic transit signals are identified using the Box-fitting Least Squares technique \citep[BLS,][]{2002A&amp;A...391..369K}.

\begin{figure*}
 \centering
 \begin{tabular}{cc}
 \includegraphics[width=9cm]{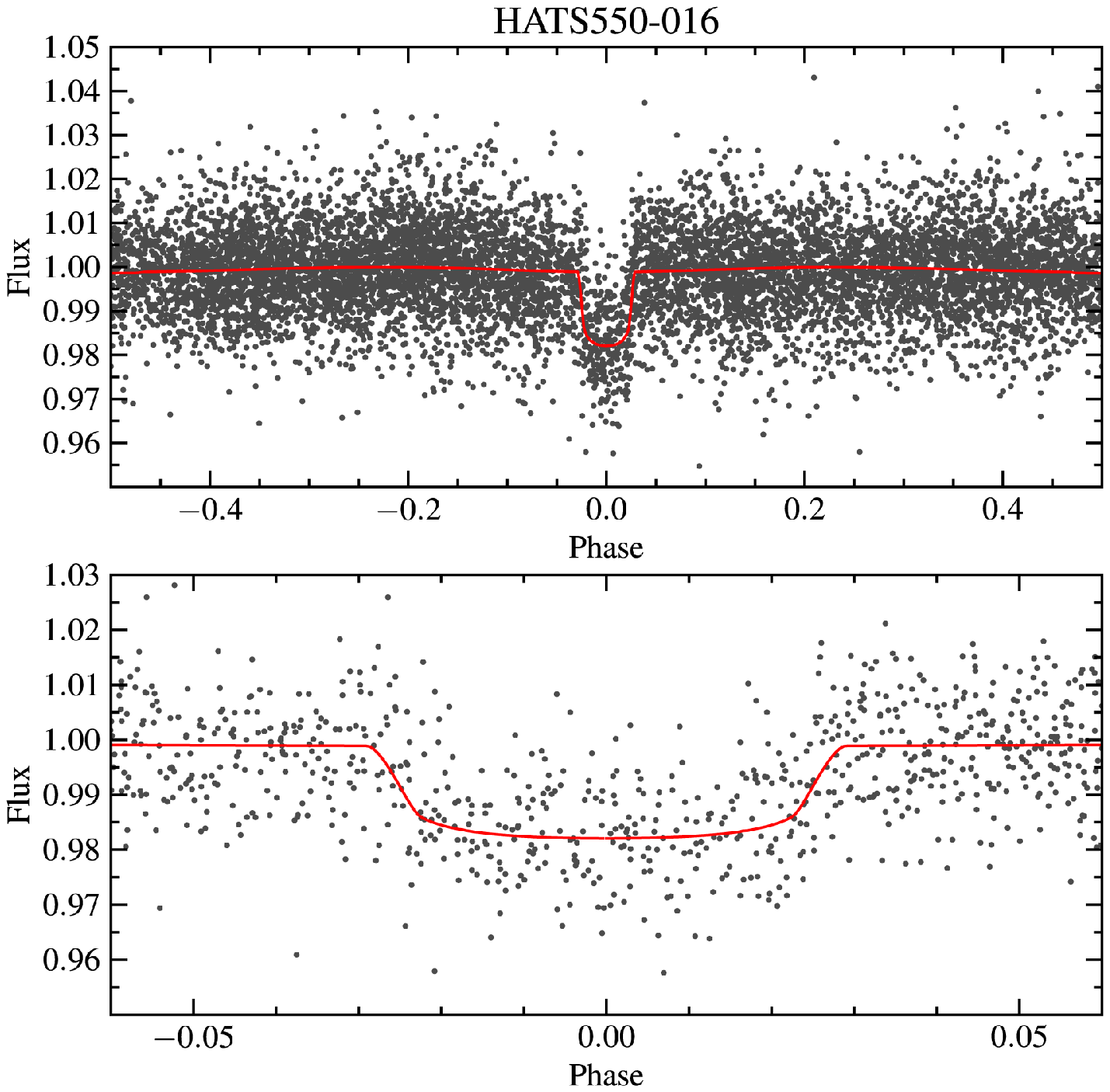} &
 \includegraphics[width=9cm]{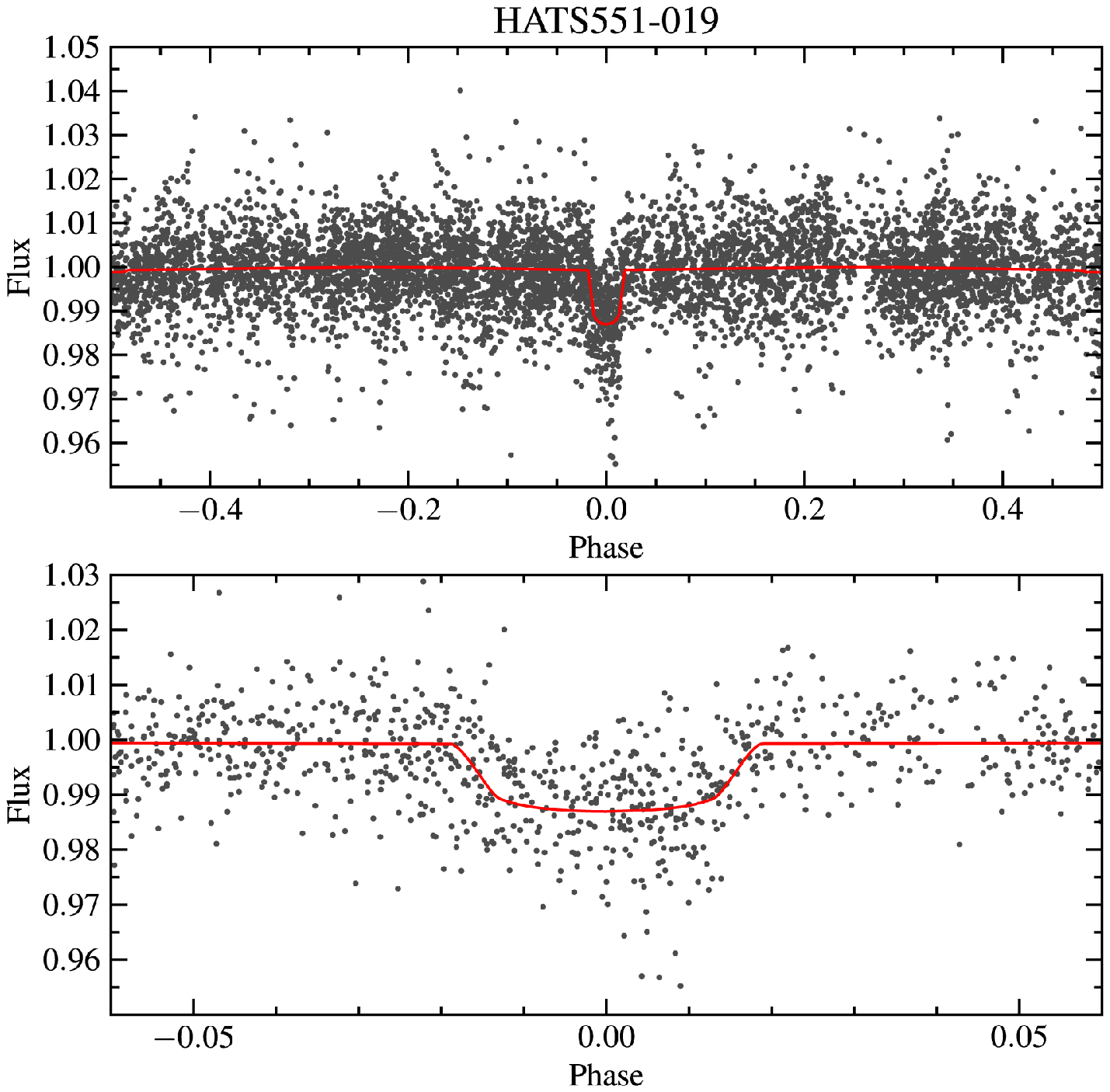} \\
 \includegraphics[width=9cm]{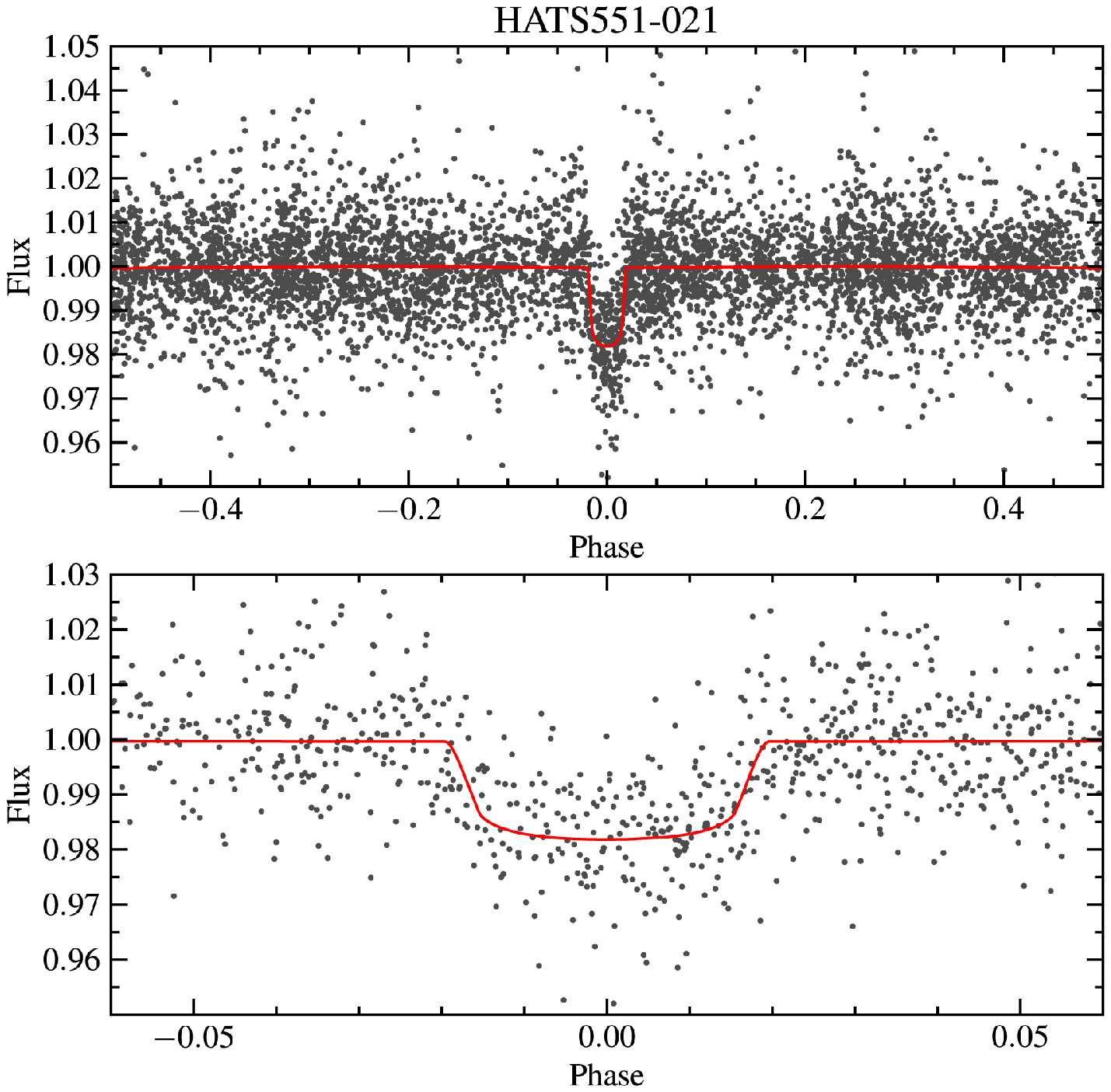} &
 \includegraphics[width=9cm]{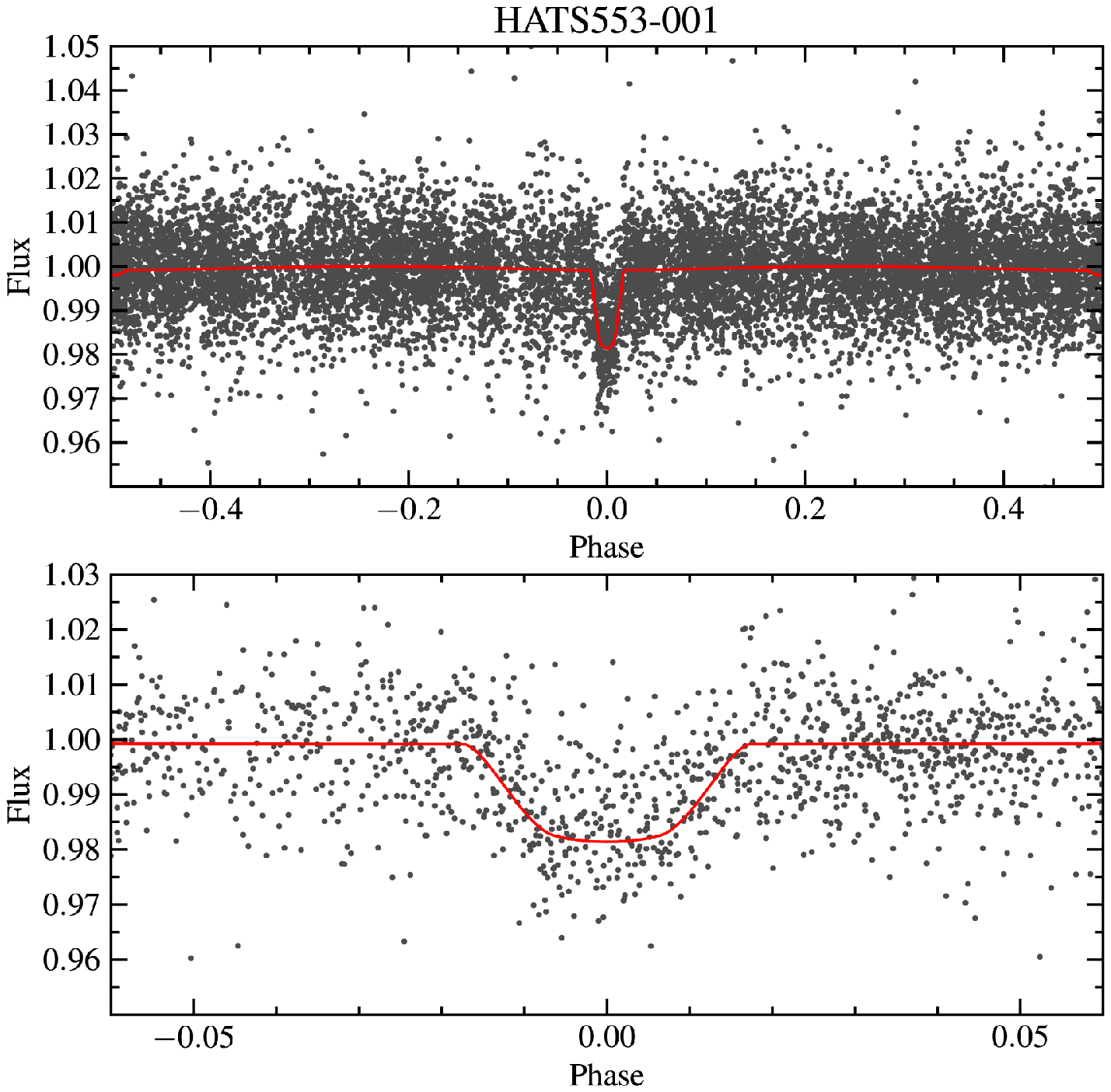} \\
 \end{tabular}
 \caption{HATSouth discovery light curves of the four VLMS systems, including close-ups of the transit event. The best fit models from Section~\ref{sec:lightcurve-analysis} are plotted in red.}
  \label{fig:HS_lightcurve}
\end{figure*}

The HATSouth discovery light curves for the systems presented in this study are shown in Figure~\ref{fig:HS_lightcurve}, and are summarised in Table~\ref{tab:phot_obs_tab}. Details of our planetary candidate selection, vetting, and confirmation process can be found in the recent HATSouth publications \citep{2013AJ....145....5P,2013A&amp;A...558A..55M,2013AJ....146..113B}.

\begin{table*}
\begin{center}
\caption{
    Summary of photometric observations
    \label{tab:phot_obs_tab}
}
{\footnotesize\begin{tabular}{lllrr}
\hline
    \multicolumn{1}{c}{Facility}          &
    \multicolumn{1}{c}{Date(s)}             &
    \multicolumn{1}{c}{Filter}      &
    \multicolumn{1}{c}{Number of images}         &
    \multicolumn{1}{c}{Cadence (s)}            \\
\hline
\bf{HATS550-016}\\
    HATSouth & 2009/09/28--2010/12/20 & $r'$ & 8726 & 240\\
    FTS / Merope & 2012/11/17 & $i'$ & 160 & 60\\
    MPG/ESO 2.2\,m / GROND & 2012/12/08 & $g',i',z'$ & 187 & 145 \\
    MPG/ESO 2.2\,m / GROND & 2012/12/08 & $r'$ & 185 & 145 \\

\\
\bf{HATS551-019}\\
    HATSouth & 2009/09/09--2010/04/29 & $r'$ & 5274 & 240\\
    PEST & 2012/12/23 & $R_c$ & 168 & 120 \\
\\
\bf{HATS551-021}\\
    HATSouth & 2009/09/09--2010/04/29 & $r'$ & 5274 & 240\\
\\
\bf{HATS553-001}\\
    HATSouth & 2009/09/17--2010/09/10 & $r'$ & 10703 & 240\\
   PEST & 2012/12/22 & $R_c$ & 92 & 120 \\
\hline
\end{tabular}}              
\end{center}   
\end{table*}

\subsection{Identification of stellar mass binaries by ANU 2.3\,m/WiFeS}
\label{sec:recon_spec}

Spectroscopic follow-up of HATSouth candidates start with reconnaissance spectroscopic observations, at high signal-to-noise (S/N) and low--medium resolution, to determine preliminary primary star properties and to search for high amplitude radial velocity variations $(>2\,\text{kms}^{-1})$. These observations allow efficient identification of non-planetary systems, such as F-M binaries, by providing an initial mass estimate for the primary star and any secondary companion found. The reconnaissance observations are summarised in Table~\ref{tab:spec_obs_tab}.

\begin{table*}
\begin{center}
\caption{
    Summary of spectroscopic observations \label{tab:spec_obs_tab}
}
{\footnotesize\begin{tabular}{llrrr}
\hline
    \multicolumn{1}{c}{Facility}          &
    \multicolumn{1}{c}{Date} &
    \multicolumn{1}{c}{Resolution} &
    \multicolumn{1}{c}{Wavelength}          &
    \multicolumn{1}{c}{Number of}\\
    \multicolumn{1}{c}{}          &
    \multicolumn{1}{c}{Range} &
    \multicolumn{1}{c}{} &
    \multicolumn{1}{c}{Coverage (\AA)}          &
    \multicolumn{1}{c}{Observations}\\
\hline
\bf{HATS550-016}\\
    ANU 2.3\,m / WiFeS & 2012/05/11--2012/08/07 & 3000 & 3500--6000 & 2 \\
    ANU 2.3\,m / WiFeS & 2012/08/04--2012/10/31  & 7000 & 5200--7000 & 16\\
    Euler 1.2\,m / Coralie & 2012/08/25--2012/11/11 & 60000 & 3850--6900 & 7\\
    ANU 2.3\,m / Echelle & 2012/12/04--2012/12/06 & 24000 & 4200--6725 & 7\\
\\
\bf{HATS551-019}\\
    Euler 1.2\,m / Coralie & 2010/10/28--2011/02/18 & 60000 & 3850--6900 & 3\\
    ANU 2.3\,m / WiFeS & 2010/11/26--2011/04/19  & 7000 & 5200--7000 & 4 \\
    ANU 2.3\,m / Echelle & 2013/03/23--2013/04/01 & 24000 & 4200--6725 & 7\\
\\
\bf{HATS551-021}\\
    Euler 1.2\,m / Coralie & 2010/10/28-2011/02/18 & 60000 & 3850--6900 & 3\\
    ANU 2.3\,m / WiFeS & 2010/11/26--2011/09/17  & 7000 & 5200--7000 & 3 \\
    MPG/ESO 2.2\,m / FEROS & 2011/09/09-2011/09/10 & 48000 & 3500--9200 & 2\\
    ANU 2.3\,m / WiFeS & 2011/09/17 & 3000 & 3500--6000 & 1 \\
    ANU 2.3 m / Echelle & 2013/03/23--2013/04/01 & 24000 & 4200--6725 & 7\\
\\
\bf{HATS553-001}\\
    ANU 2.3\,m / WiFeS & 2012/05/08 & 3000 & 3500--6000 & 1 \\
    ANU 2.3\,m / WiFeS & 2012/05/09--2012/05/09  & 7000 & 5200--7000 & 2\\
    ANU 2.3\,m / Echelle & 2013/03/23--2013/04/01 & 24000 & 4200--6725 & 7\\
\hline
\end{tabular}}
\end{center}
\end{table*}

Initial spectroscopic observations of the targets were obtained using the Wide Field Spectrograph (WiFeS) on the ANU 2.3\,m telescope \citep{2007Ap&amp;SS.310..255D}, located at Siding Spring Observatory (SSO), Australia.

First, a low resolution ($R \equiv \lambda / \Delta \lambda = 3000$) spectrum covering the wavelength region 3500--6000\,\AA\, was used to obtain an initial stellar classification of the target star. The flux calibrated spectrum was fitted to a grid of synthetic spectra from \citet{2008A&amp;A...486..951G}. Details of the low resolution spectral reduction and analysis, including fitting of the stellar properties, are given in \citet{2013AJ....146..113B}. These stellar parameters are later refined using high resolution spectra (see Section~\ref{sec:host-star-properties}).

Subsequent medium resolution, multi-epoch radial velocity observations were performed with WiFeS at $R=7000$, observed at phase quadratures of the photometric ephemeris, where the potential velocity variation is greatest. In the case of stellar binaries, the radial velocity variations are often apparent with two well-time exposures. Combined with the WiFeS stellar parameters, the WiFeS radial velocity orbit provides an initial mass estimate of the companions and affect their prioritisation for further follow-up studies. The WiFeS velocities were not included in the final system analysis, since fewer, lower resolution observations do not contribute greatly to improving the precision of the measured radial velocity orbit.

\subsection{High resolution spectroscopic follow-up}
\label{sec:high-resol-flup}

Radial velocity measurements derived from high resolution spectroscopic observations were obtained for the VLMS systems using the Echelle spectrograph on the ANU 2.3\,m telescope at SSO; the fibre-fed echelle spectrograph CORALIE on the Swiss Leonard Euler 1.2-m telescope \citep{2000A&amp;A...354...99Q} at La Silla Observatory (LSO), Chile, and the fibre-fed echelle spectrograph FEROS on the MPG/ESO 2.2\,m telescope \citep{1998SPIE.3355..844K} at LSO. The observations are summarised in Table~\ref{tab:spec_obs_tab}. Descriptions of the data reduction and analysis for CORALIE and FEROS can be found in \citet[][Jord\'{a}n et al. in prep]{2013AJ....145....5P}. This is the first time we have used the ANU 2.3\,m Echelle to monitor HATSouth targets, a description of these observations are presented below. 

\subsubsection{ANU 2.3\,m / Echelle}
\label{sec:echelle}

High resolution spectra of the systems were obtained using the Echelle spectrograph on the ANU 2.3\,m telescope. The echelle was configured to a 1\farcs8 wide slit, delivering a resolution of $ R = 24000$, velocity dispersion of $4.0 \, \text{km} \, \text{s}^{-1} \, \text{pixel}^{-1}$, in the spectral range 4200--6725\,\AA, over 20 echelle orders. The detector is a $2\text{K}\times 2\text{K}$ CCD with gain of $2\,\text{e}^-\,\text{ADU}^{-1}$ and read noise of $2.3 \,\text{ADU}\,\text{pixel}^{-1}$, and binned $2\times$ in the spatial direction. A number of instrument limitations prevent us from achieving better than $500\,\text{ms}^{-1}$ velocity precision. For example, the instrument is mounted on the Nasmyth focus, not in a temperature stabilised environment; the low efficiency of the spectrograph limits the study to only bright stars $(<13.5\,V_\text{mag})$. The data was reduced with the IRAF\iraf package CCDPROC, and extracted using ECHELLE\@. A rapidly rotating B star spectrum is divided through each observation to remove the blaze function. A low order spline interpolation is then used to continuum normalise each spectrum. The wavelength solution was provided by Th-Ar arc lamp exposures that bracketed each science exposure. 

Radial velocity measurements were obtained by cross correlating the object spectra against a series of radial velocity standard star spectra taken on the same night. The radial velocities derived from selected echelle orders not severely contaminated by telluric absorption features were sigma clipped and weight averaged according to their respective CCF heights and S/Ns. Typically 15 echelle orders were used in the cross correlations. A velocity and the associated uncertainty was determined for each order, from which the weighted average and standard deviation were calculated and adopted as the measured velocity. For stable HATSouth candidates with $V_\text{mag}\approx 13$, the long-term root-mean-squared (RMS) velocity scatter of the instrument is $\sim 1.0\,\text{km}\,\text{s}^{-1}$. Stellar parameters were also derived from the Echelle spectra, the process is described in detail in Section~\ref{sec:host-star-properties}. 

\subsection{Photometric follow-up}
\label{sec:phot-flup}

Follow-up photometric confirmations for the transit events of HATS550-016B, HATS551-019B, and HATS553-001B were made using the Merope camera on the 2\,m Faulkes Telescope South (FTS) located at SSO, the Gamma-Ray Burst Optical/Near-Infrared Detector (GROND) on the MPG/ESO 2.2\,m telescope at LSO, and the 0.30\,m Perth Exoplanet Survey Telescope (PEST) located in Perth, Australia. The observations are listed in Table~\ref{tab:phot_obs_tab}, with the light curves plotted in Figures~\ref{fig:fit_550-016}, \ref{fig:fit_551-019}, \ref{fig:fit_553-001}.

\subsubsection{FTS 2\,m / Merope}
\label{sec:fts-phot}

A near-full transit for HATS550-016B was observed on 2012 November 17 using the Merope imaging camera on the 2\,m Faulkes Telescope South, part of the Las Cumbres Global Telescope (LCOGT) Network. Merope is a $2\text{K} \times 2\text{K}$ camera with $0.139\text{"}\,\text{pixel}^{-1}$ pixels, binned at $2\times2$, and a FoV of $4.7\text{'}\times4.7\text{'}$. The observation was performed in the SDSS $i'$ band, with 60s exposure time, and the telescope slightly defocused to reduce pixel-pixel and imperfect flat field systematic effects and to prevent saturation. The data was reduced by the automated LCOGT pipeline.

Aperture photometry was performed on the reduced images with Source Extractor \citep{1996A&amp;AS..117..393B}. Stellar flux was extracted over multiple diameter apertures, reference stars are selected based on their brightness, colour, and lack of blended neighbours. 

\subsubsection{MPG/ESO 2.2\,m / GROND}
\label{sec:grond}

A full transit for HATS550-016B was also observed on 2012 December 08 using GROND on the 2.2\,m MPG/ESO telescope at LSO. GROND provides simultaneous multi-band photometry in four optical bands similar to Sloan $g'$, $r'$, $i'$, and $z'$. It has a field of view of $5.4\text{'} \times 5.4\text{'}$ with $0.158\text{"}\,\text{pixel}^{-1}$ pixels. Exposure times for the HATS550-016 observations were 100s. Details of the GROND observing strategy and data reduction procedure can be found in \citet{2013AJ....145....5P,2013A&amp;A...558A..55M}.

\subsubsection{PEST}
\label{sec:tg30cm}

A full transit of HATS553-001 and a partial transit of HATS551-019 were observed using PEST on 2012 December 22 and 2012 December 23 respectively. PEST is a fully automated 0.30\,m Meade LX200 Schmidt Cassegrain telescope located at latitude $-31^\circ\, 59\text{'} \, 34\text{''}$, longitude $115^\circ \, 47\text{'} \, 53\text{''}$E. The telescope is coupled with a focal reducer to yield a focal ratio of f/5. The detector is a SBIG ST-8XME CCD camera with gain of $2.27\,e^{-}\,\text{ADU}^{-1}$ and read noise of $19.9\,e^{-}$, with image scale of $1.2\text{''}\,\text{pixel}^{-1}$, and a FoV of $31\text{'} \times 21 \text{'}$. Images were taken in the $Rc$ band, with the telescope in focus, exposure times are provided in Table~\ref{tab:phot_obs_tab}. Typical conditions yield stellar point-spread-functions with FWHM of $\sim 3$ pixels. Flat field frames are taken in twilight whenever possible. Dark frames of equal exposure length to the object frames are drawn from a library of master dark frames, renewed every month. 

Image reduction and aperture photometry were performed using the C-Munipack program. Relative photometry is performed, with the reference light curve made from the weighted average of high S/N field star light curves.

\section{Analysis}
\label{sec:analysis}

\subsection{Fundamental stellar atmospheric parameters}
\label{sec:host-star-properties}

The fundamental primary star properties, including effective temperature (\teff), surface gravity (\logg), metallicity (\feh), and projected rotational velocity (\vsini), were derived by fitting synthetic model spectra to the averaged ANU 2.3\,m / Echelle observations. We generated a synthetic spectral library with the ATLAS9 model atmospheres \citep{2004astro.ph..5087C}, using the spectral synthesis program SPECTRUM\footnote{\texttt{http://www1.appstate.edu/dept/physics/spectrum/spectrum.html}} \citep{1994AJ....107..742G}. The spectra have resolutions of $R=24000$, matching that of the Echelle instrument, and were generated at the ATLAS9 \teff, \logg, and \feh grid points, using the default isotopic line lists provided with SPECTRUM, then broadened to multiple rotational velocities spaced 5\,\kms apart. Model microturbulences are fixed at 2\,\kms, given that the range of possible microturbulence values for the \teff and \logg tested varies only by $\sim 0.5\,\text{km}\,{s}^{-1}$ \citep{2013A&amp;A...553A...6H}. The individual echelle orders of the observed spectrum were matched to a restricted grid, centred about the rough stellar parameter estimates from the WiFeS spectrum. Using exposures of standard stars, we found echelle orders that gave the most reliable stellar parameters for a target of a given spectral type. The fitting results from these selected orders were weight averaged according to their RMS scatter from the fit.     The \logg was subsequently constrained from theoretical isochrones (see Section~\ref{sec:lightcurve-analysis}) via global light curve and radial velocity modelling. We then performed the grid search again, at a finer \vsini grid of 1\,\kms spacings, with \logg fixed, to obtain the final stellar parameters. The primary star parameters are presented in Table~\ref{tab:host-star}.

\begin{table*}
\begin{center}
\caption{
    Primary star properties \label{tab:host-star}
}
{\footnotesize\begin{tabular}{lrrrr}
\hline
    \multicolumn{1}{c}{Property}          &
    \multicolumn{1}{c}{HATS550-016} &
    \multicolumn{1}{c}{HATS551-019} &
    \multicolumn{1}{c}{HATS551-021} &
    \multicolumn{1}{c}{HATS553-001} \\
\hline
GSC$^{a}$ & 6465-00602 & 6493-00290 & 6493-00315 & 5946-00892 \\
RA$^{b}$ (J2000 HH:MM:SS.SS) & 04:48:23.32 & 05:40:46.16 & 05:42:49.12 &06:16:00.66 \\
DEC$^{b}$ (J2000 DD:MM:SS.SS) & -24:50:16.88 & -24:55:35.16 & -25:59:47.49 & -21:15:23.82\\
\\
\multicolumn{4}{l}{\bf{Photometric Properties}}\\
$V$$^{c}$ & $13.605\pm0.041$ & $12.058\pm0.006$ & $13.114\pm0.008$ & $13.189\pm0.021$\\
$B$$^{c}$ & $14.052\pm0.020$ & 12.497 & $13.580\pm0.012$ & $13.694\pm0.014$\\
$g'$$^{c}$ & $13.782\pm0.032$ & 12.198 & 13.308 & 13.408\\
$r'$$^{c}$ & $13.499\pm0.031$ & 11.953 & $13.044$ & 13.102\\
$i'$$^{c}$ & 13.438 & $11.922\pm0.038$ & $12.985\pm0.077$ & 13.988\\
$J$$^{b}$ & $12.640\pm0.021$ & $11.146\pm0.021$ & $12.150\pm0.021$ & $12.245\pm0.021$\\
$H$$^{b}$ & $12.379\pm0.025$ & $10.943\pm0.024$ & $11.956\pm0.020$ & $12.023\pm0.025$ \\
$K$$^{b}$ & $12.308\pm0.021$ & $10.914\pm0.021$ & $11.873\pm0.023$ & $11.970\pm0.024$ \\
\\
\multicolumn{4}{l}{\bf{Derived Spectroscopic Properties}}\\
\teff\,(K) & $6420\pm90$ & $6380\pm170$ & $6670 \pm 220$ & $6230\pm250$ \\
\feh & $-0.60\pm0.06$ & $-0.4 \pm 0.1$ & $-0.4 \pm 0.1$ & $-0.1 \pm 0.2$ \\
\vsini\,(\kms) & $30.0\pm1.7$ & $17.1 \pm 2.0$ & $16.4 \pm 10.6$ & $22.2\pm 1.8$ \\
\hline
\end{tabular}}
\end{center}

\begin{flushleft} 
$^{a}${Hubble guide star catalogue}\\
$^{b}${2MASS}\\
$^{c}${APASS Data Release 7, uncertainties are quoted where available as the scatter from multiple observations.}
\end{flushleft}
\end{table*}

To investigate the errors in our spectral typing pipeline, we observed seven reference stars from \citet{2005ApJS..159..141V}, and four known planet hosting stars of similar spectral type and brightness to our candidates \citep[WASP-61, 62, 78, 79][]{2012MNRAS.426..739H,2012A&amp;A...547A..61S}. These observations were used to determine the echelle orders that yielded the most reliable spectral types. Reference stars with $v \sin i > 10\,\text{km}\,\text{s}^{-1}$ were also analysed at the finer 1\,\kms \vsini grid spacing. We are not sensitive to $v\sin i < 10\,$\kms rotational velocities, where we become limited by the instrument resolution. The RMS difference between our measured stellar parameters and literature values are 113\,K in \teff, 0.19\,dex in \logg, 0.12\,dex in \feh, and 4.4\,\kms in \vsini. After correcting for an empirical offset in each parameter, we get errors of 88\,K in \teff, 0.13\,dex in \logg, 0.09\,dex in \feh, and 0.7\,\kms in \vsini. The offsets have been incorporated in the stellar parameters presented here.

Macroturbulence and microturbulence are free parameters in 1D stellar atmosphere models that contribute to the overall broadening of the spectral features. We investigate the degeneracy between these parameters and the measured stellar rotational broadening. The spectrum of HATS550-016 was fitted to synthetic spectral grids generated at macro and microturbulences from 0 to 5\,\kms, producing a mean variation of 0.8\,\kms in the resulting \vsini measurement, smaller than our quoted measurement errors. Since the rotational broadening parameter for all host stars studied are much larger than the broadening from macro and microturbulence, we conclude minimal systematic uncertainty contributions from these parameters. In addition, we checked for the dependency of the measured \vsini to the synthetic spectral resolution, and found a variation of 0.6\,\kms over a change in resolution of 1000. Since the resolution is measured from the ThAr arc lamp spectra, we expect no significant contribution from uncertainties in the resolution to the \vsini systematic uncertainties.

\subsection{Global modelling of data}
\label{sec:lightcurve-analysis}

To derive the system properties, we performed simultaneous modelling of the HATSouth discovery photometry, follow-up photometry where available, and radial velocity measurements. The light curves were modelled using the \citet{1972ApJ...174..617N} model for eclipsing binaries, allowing for ellipsoidal phase variations, as implemented in the JKTEBOP code \citep{1981AJ.....86..102P,2004MNRAS.351.1277S}. The Keplerian orbit was used to model the radial velocity measurements. 

The free parameters in the global fit include the orbital period $P$, transit time $T_0$, radius ratio $R_2/R_1$, normalised radius sum $R_\text{sum}=(R_1+R_2)/a$, radial velocity orbit semi-amplitude $K$, eccentricity parameters $e\cos\omega$ and $e\sin\omega$, and inclination $i$. Quadratic limb darkening coefficients for the primary star were fixed to values from \citet{2000A&amp;A...363.1081C}. We assume uniform priors for all the free parameters. In the cases where the proposed iteration in inclination is $i>90^\circ$, we adopt the $180^\circ-i$ geometry to avoid the discontinuous boundary. To account for the non-zero flux contribution from the M-dwarf companion, we also obtained an approximate surface brightness estimate for the companion using a 5 Gyr \citet{1998A&amp;A...337..403B} isochrone. However, the flux contribution from the M-dwarf is $\ll 0.1\text{\%}$ in the $R$ band and is negligible. Ellipsoidal variations are included in the model by including the mass ratio parameter $q$, using masses determined per iteration from isochrone fitting (see Section~\ref{sec:isochrone-fitting}). The best fitting parameters and the surrounding error space were explored by the \emph{emcee} \citep{2013PASP..125..306F} implementation of an Markov chain Monte Carlo (MCMC) ensemble sampler, with the individual measurement errors for all datasets (discovery, follow-up photometry, radial velocity) inflated such that the reduced \chisq is at unity.

Instrumental offsets were derived for each instrument separately. In addition, the HATSouth discovery light curves can be diluted in eclipse depth if they are treated by the TFA detrending algorithm. In the cases where well sampled follow-up photometry is available (HATS550-016 and HATS553-001), we fitted for a dilution factor for the HATSouth light curves, using the follow-up light curve as reference. Where follow-up photometry of the full transit is not available (HATS551-019 and HATS551-021), simultaneous TFA corrections were performed on the HATSouth light curve using the transit model for each MCMC iteration \citep[see Section 6,][]{2005MNRAS.356..557K}.

\subsection{Determination of mass and radius }
\label{sec:determ-mass-radi}

We derive the mass and radius of the primary and secondary stars at each iteration via 1) determination of the primary star properties from stellar isochrones using measured spectroscopic and light curve parameters, 2) assuming spin-orbit synchronisation for the system, and derive mass and radius from the spectroscopic \vsini measurement, and 3) a combined analysis that includes isochrone fitting and the assumption of spin-orbit synchronisation. Each analysis employs a separate MCMC routine that explores their respective posteriors. 


\subsubsection{Isochrone fitting}
\label{sec:isochrone-fitting}
\citet{2007ApJ...664.1190S} showed that the normalised orbital distance $a/R_1$, derived from the global fit, can be combined with model isochrones to refine the stellar atmosphere parameters. $a/R_1$ is related to the mean stellar density by
\begin{equation}
  \label{eq:aRs_eq}
  \frac{M_1}{R_1^3} = \frac{4\pi^2}{G P^2} \left(\frac{a}{R_1} \right)^3 - \frac{M_2}{R_1^3} \, .
\end{equation}
Although the second term is usually insignificant and often discarded in the case of transiting planets, it cannot be ignored for stellar mass companions \citep{2013Aamp;A...549A..18T}. Therefore, for each iteration of the minimisation and MCMC routines, we used initial estimates of $M_1$ and $R_1$ to derive $M_2$ from the radial velocity orbit, then used the $M_2$ estimate and the fitted $a/R_1$, and spectroscopically determined \teff and \feh to derive new theoretical $M_1$ and $R_1$ values by isochrone fitting. Finally, the new primary mass and radius were used to derive a refined $M_2$ estimate. The Yonsei-Yale isochrones \citep{2001ApJS..136..417Y} provide the theoretical stellar masses and radii. To incorporate the uncertainties in the spectroscopic stellar parameters into the error analysis, \teff and \feh were drawn from Gaussian distributions in the MCMC routine. This method also gives us a more precise estimate for \logg of the primary star. This \logg value was incorporated into the spectral classifications process (Section~\ref{sec:host-star-properties}) to better constrain the \teff, \feh, and \vsini estimates.


\subsubsection{Spin-orbit synchronisation}
\label{sec:spin-orbit-synchr}
For stellar mass binaries at short periods, spin-orbit synchronisation via tidal interactions is expected to occur within $\sim 100$\,Myr \citep{1977A&amp;A....57..383Z,1981A&amp;A....99..126H}. If we assume spin-orbit synchronisation for these systems, and that the stellar spin-axis is near-perpendicular to our line-of-sight $(i_\text{orb} = i_\text{rot})$, then it is also possible to derive model-independent estimates of the stellar masses and radii \citep[e.g.][]{2007ApJ...663..573B}. The masses and radii of components A and B can be calculated from purely observable quantities using:
\begin{align}
  \label{eq:sync-mass-radius}
  M_1 &= \frac{P}{2\pi G} \left(\frac{a}{R_1}\right)^2 (v\sin i_\text{rot})^2 \\ & \times \left[ \frac{(a/R_1) v \sin i_\text{rot} - K\sqrt{1-e^2}}{\sin^3 i_\text{orb}} \right] \nonumber \\
  M_2 &= \frac{P}{2\pi G}  \left(\frac{a}{R_1}\right)^2 \left[ \frac{K (v\sin i_\text{rot})^2 \sqrt{1-e^2}}{\sin^3 i_\text{orb}} \right] \\
  R_1 &= \frac{P}{2\pi} \frac{v \sin i_\text{rot}}{\sin i_\text{orb}} \\
  R_2 &= R_1 \left(\frac{R_2}{R_1}\right) \,.
\end{align}

We caution that spin-orbit synchronisation and the alignment of the stellar spin-axis should not be automatically assumed. Although rapid synchronisation is expected for binary systems, some previous F-M binaries have been measured to be asynchronous \citep[e.g.][]{2005A&amp;A...433L..21P,2006A&amp;A...447.1035P,2013Aamp;A...549A..18T}. In addition, whilst the alignment of the stellar spin and companion orbital axes is also often predicted from formation and tidal interactions, (and by extension of the transit geometry the stellar spin-axis should also be perpendicular to our line-of-sight) spin-orbit misaligned stellar binaries have been identified \citep[DI Her, KOI-368,][]{2009Natur.461..373A,2013ApJ...776L..35Z}. It is therefore necessary to compare the stellar parameter results from synchronisation against that of isochrone fitting before this method can be adopted.

For each iteration of the global minimisation and MCMC routine, we also calculated the primary and companion masses and radii assuming synchronisation. The adopted \vsini value was given by the spectroscopic analysis, and was drawn from a Gaussian distribution in the MCMC error analysis.

\subsubsection{Combined mass-radius estimate}
\label{sec:combined-mass-radius}

We also perform a combined analysis, where the mass and radius are calculated using isochrone fitting as described in Section~\ref{sec:isochrone-fitting}. The expected \vsini is then derived using the period and radius, and compared to the measured \vsini. We calculate an additional \chisq term,
\begin{equation}
  \label{eq:chisq-vsini}
  \chi^2_{v\sin i_\text{rot}} = \left(\frac{2\pi R_1 \sin i_\text{orb}/P - v \sin i_\text{rot}}{\Delta v \sin i_\text{rot}}\right)^2 \, ,
\end{equation}
which is added to the \chisq from the light curve and radial velocity data. Due to the transit geometry, we approximate $\sin i_\text{orb}\approx 1$ in the calculation. The MCMC minimisation is re-run for this combined analysis.\\

\begin{figure*}
  \centering

  \begin{tabular}{cc}
      \includegraphics[width=9cm]{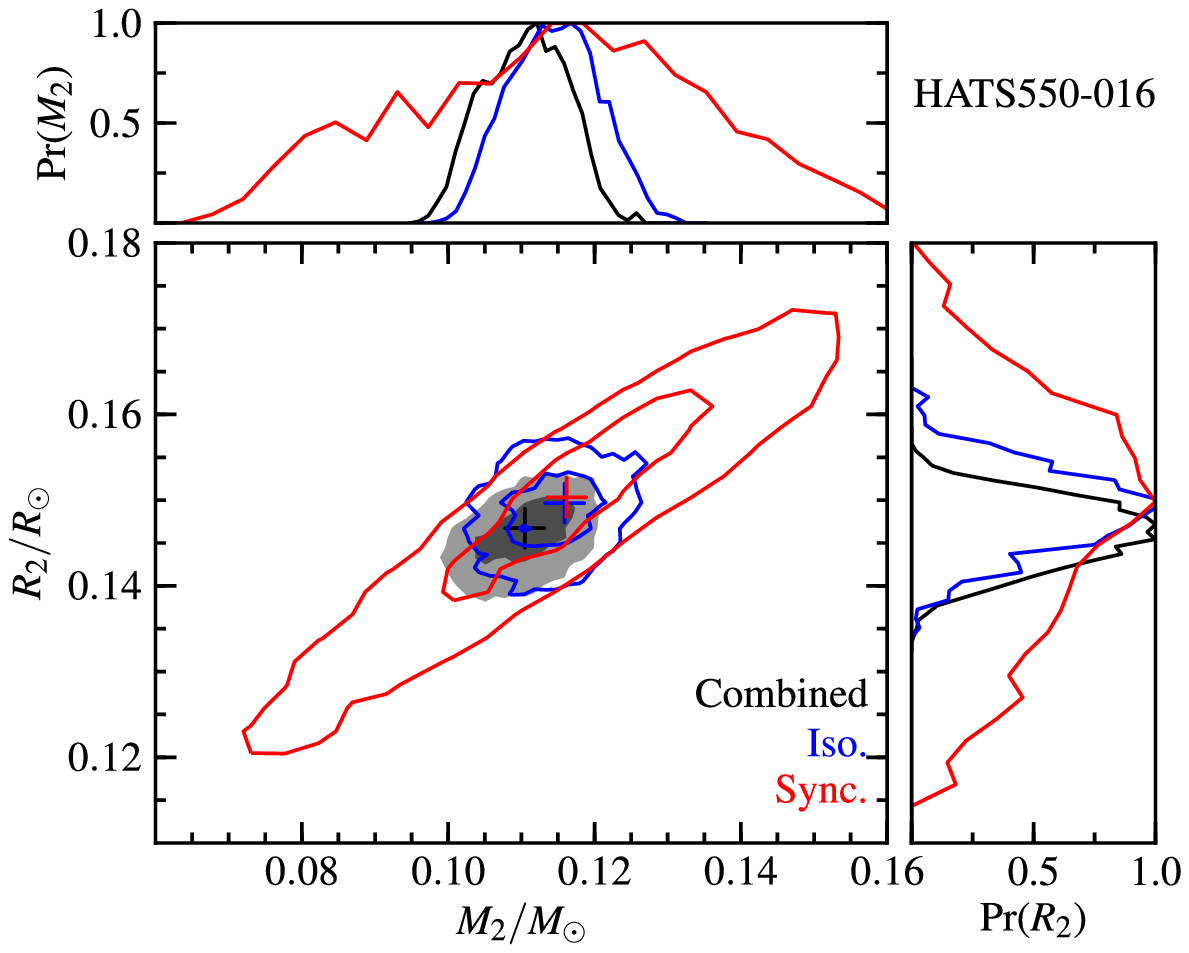}&
      \includegraphics[width=9cm]{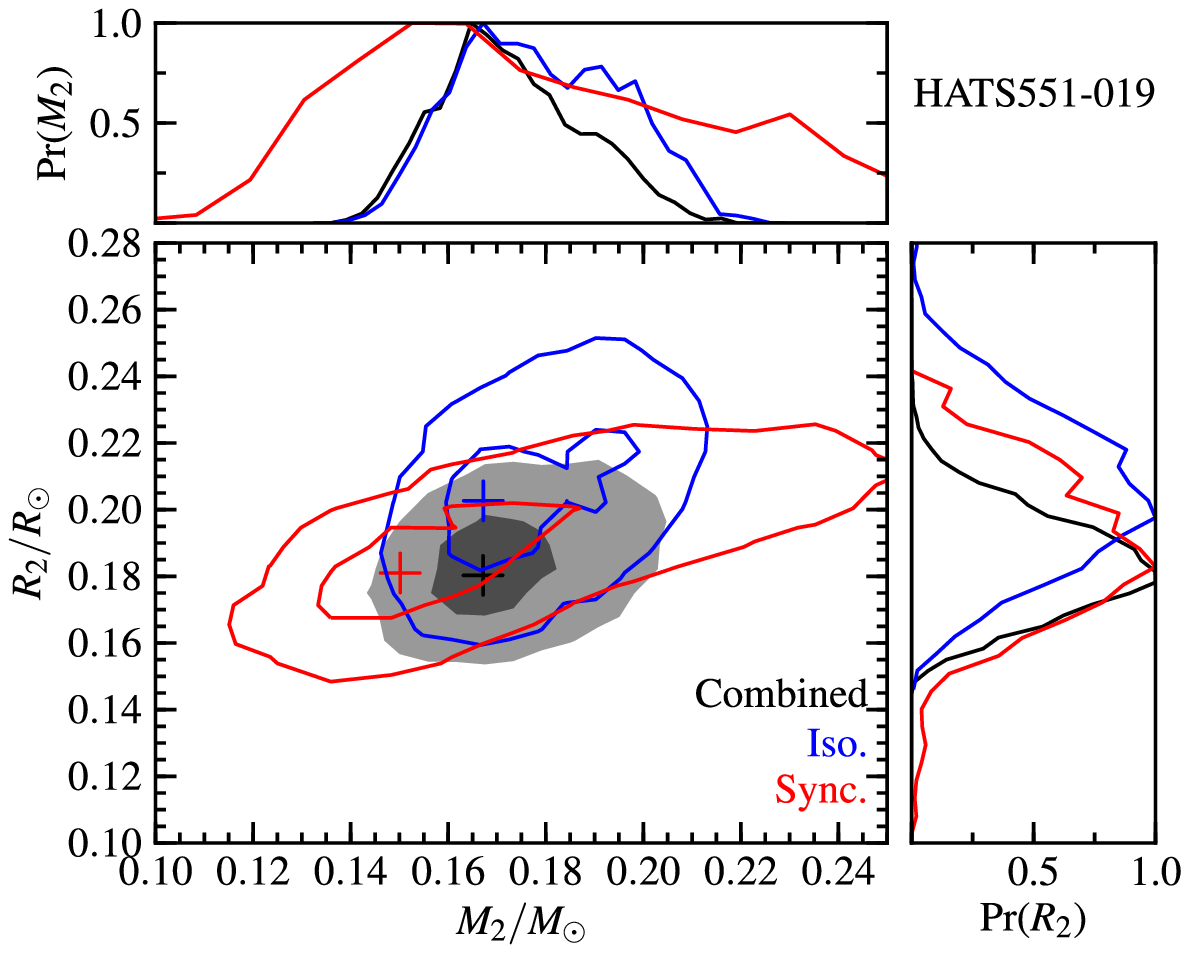}\\
      \includegraphics[width=9cm]{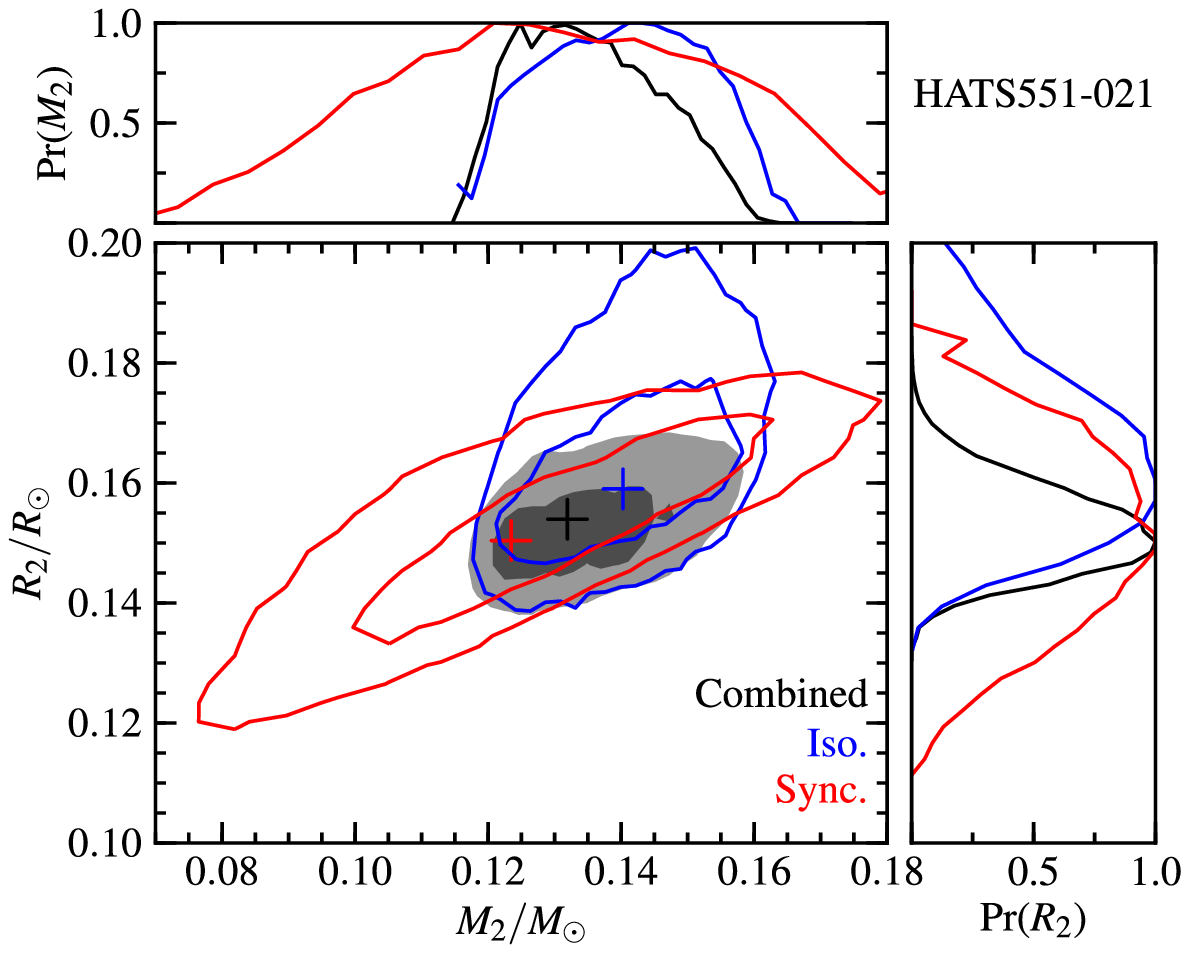}&
      \includegraphics[width=9cm]{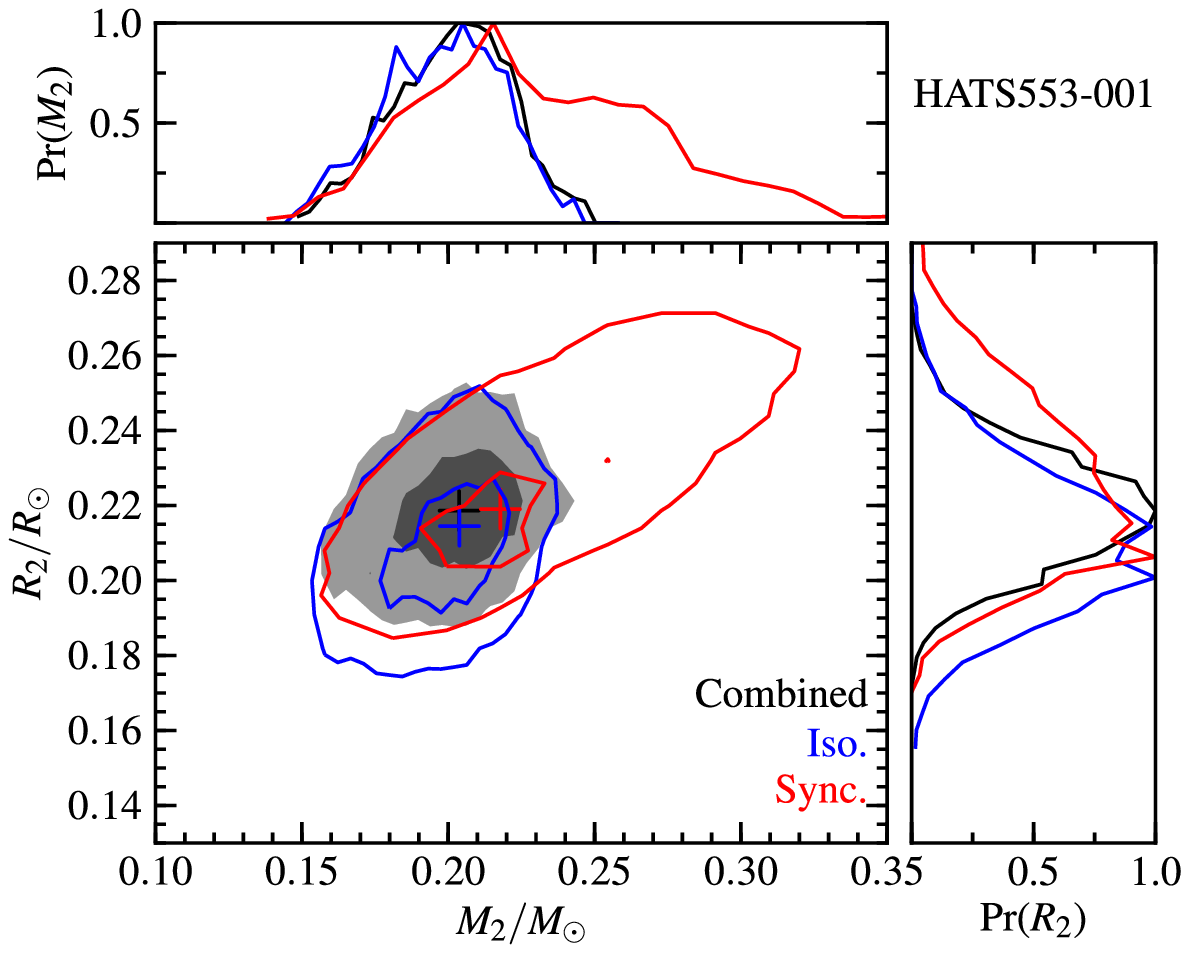}\\
  \end{tabular}
  \caption{The 1 and 2$\sigma$ confidence regions for the masses and radii of the VLMS systems presented in this study. We plot individual confidence regions derived from the isochrone fit (blue), assuming spin-orbit synchronisation (red), and the combined analysis (gray). The confidence regions from the combined analysis is adopted. The corresponding crosses mark the peak of the probability distributions.}
  \label{fig:posterior}
\end{figure*}

The probability distributions for the mass and radius, measured using the above techniques, are plotted in Figure~\ref{fig:posterior}. We find that the $2\sigma$ confidence regions derived using isochrone and synchronisation overlap for all the systems. This indicates that the assumption of spin-orbit synchronisation, required for the combined analysis (Section~\ref{sec:combined-mass-radius}), is valid for all systems. The system parameters from the combined analysis is adopted for discussion here onwards. The stellar and system parameters are presented in Table~\ref{tab:stellar-params}.

\input{params_table.tex}

\input{system_plots.tex}

We also tested the sensitivity of the final results against various assumptions in the methods outlined above. These tests were performed on the HATS550-016 dataset. The limb darkening coefficients were set free and allowed to vary within 0.2 of the values from \citet{2000A&amp;A...363.1081C}. The resulting system parameters did not deviate from those presented in Table~\ref{tab:stellar-params}, with no significant increase in the uncertainties. To test the dependence of the results on the Yonsei-Yale isochrones, we performed the analysis in Section~\ref{sec:isochrone-fitting} using the Dartmouth isochrones \citep{2008ApJS..178...89D}, but found no deviation in the results. To test the effectiveness of reconstructive TFA  \citep{2005MNRAS.356..557K} at recovering the true transit shape, we analysed the HATS550-016 system using only the HATSouth discovery data by excluding the follow-up photometry observations, and found no significant deviation in the results, however the uncertainties were increased by a factor of $\sim2-3$. We caution that correlated noise, present in the follow-up light curve, were unaccounted for in the analysis, and may lead to under-estimated uncertainties.

\section{Discussion}
\label{sec:discussion}

We have presented the discovery of four transiting VLMSs, with masses ranging from 0.1 to 0.2\,$M_\odot$. Their properties are discussed briefly below. 

\subsection{HATS550-016B}
\label{sec:hats550-016b}

HATS550-016B is the lowest mass star within our sample, and is the second lowest mass star known with mass and radius determined to better than $10\text{\%}$ \citep[after J1219-39b;][]{2013Aamp;A...549A..18T}. It has a mass and radius of $0.110_{-0.006}^{+0.005}\, M_\odot$ and $0.147_{-0.004}^{+0.003}\,R_\odot$, and orbits a relatively metal deficient $(\text{[Fe/H]}=-0.60\pm0.06)$ F type star of age $5_{-1}^{+3}$ Gyr in a period of $2.051811_{-0.000002}^{+0.000002}$ days. The radius of HATS550-016B is inflated by 13\% compared to \citet{1998A&amp;A...337..403B} models, assuming it has the same metallicity as the primary star. It is the only star in this sample that is inflated with respect to the isochrones. 

\subsection{HATS551-019B}
\label{sec:hats551-019b}

HATS551-019B is a $0.17_{-0.01}^{+0.01}\,M_\odot$ , $0.18_{-0.01}^{+0.01}\,R_\odot$ VLMS with a $4.68681_{-0.00001}^{+0.00002}$ day period orbit a $6_{-2}^{+2}$ Gyr F subgiant, with sub-solar metallicity of $\text{[Fe/H]} = -0.4\pm 0.1$. Since the follow-up photometry of HATS551-019 covers only the egress event, the HATSouth discovery light curves, detrended using simultaneous TFA, are also used to constrain the $R_2/R_1$ ratio. The radius of HATS551-019B agrees with theoretical predictions to within errors.

\subsection{HATS551-021B}
\label{sec:hats551-021b}

HATS551-021B is a $0.132_{-0.005}^{+0.014}\,M_\odot$, $0.154_{-0.008}^{+0.006}\,R_\odot$ VLMS in a $3.63637_{-0.00005}^{+0.00005}$ day period orbit about an F dwarf with metallicity of $\text{[Fe/H]} = -0.4\pm0.1$. The age of the primary star is ill defined from isochrone fitting ($4_{-4}^{+3}$ Gyr). We find no chromospheric Ca HK emission, nor excess Li absorption, indicating that the system is likely $>1$ Gyr in age. No follow-up photometry is available for HATS551-021, so the $R_2/R_1$ ratio is constrained purely from HATSouth discovery light curves. The radius of HATS551-021B matches theoretical isochrones very well. 

\subsection{HATS553-001B}
\label{sec:hats553-001b}

HATS553-001B is a VLMS with mass of $0.20_{-0.02}^{+0.01}\,M_\odot$, and radius of $0.22_{-0.01}^{+0.01}\,R_\odot$, orbiting a $3_{-2}^{+2}$ Gyr F type star of near solar metallicity $(\text{[Fe/H]} = -0.1\pm0.2)$ in a $3.80405_{-0.00001}^{+0.00001}$ day period orbit. The radius of HATS553-001B matches the isochrones to within errors.

\subsection{Mass--radius relationship}
\label{sec:mass-radi-relat}

\begin{figure*}
  \centering
  \includegraphics[width=15cm]{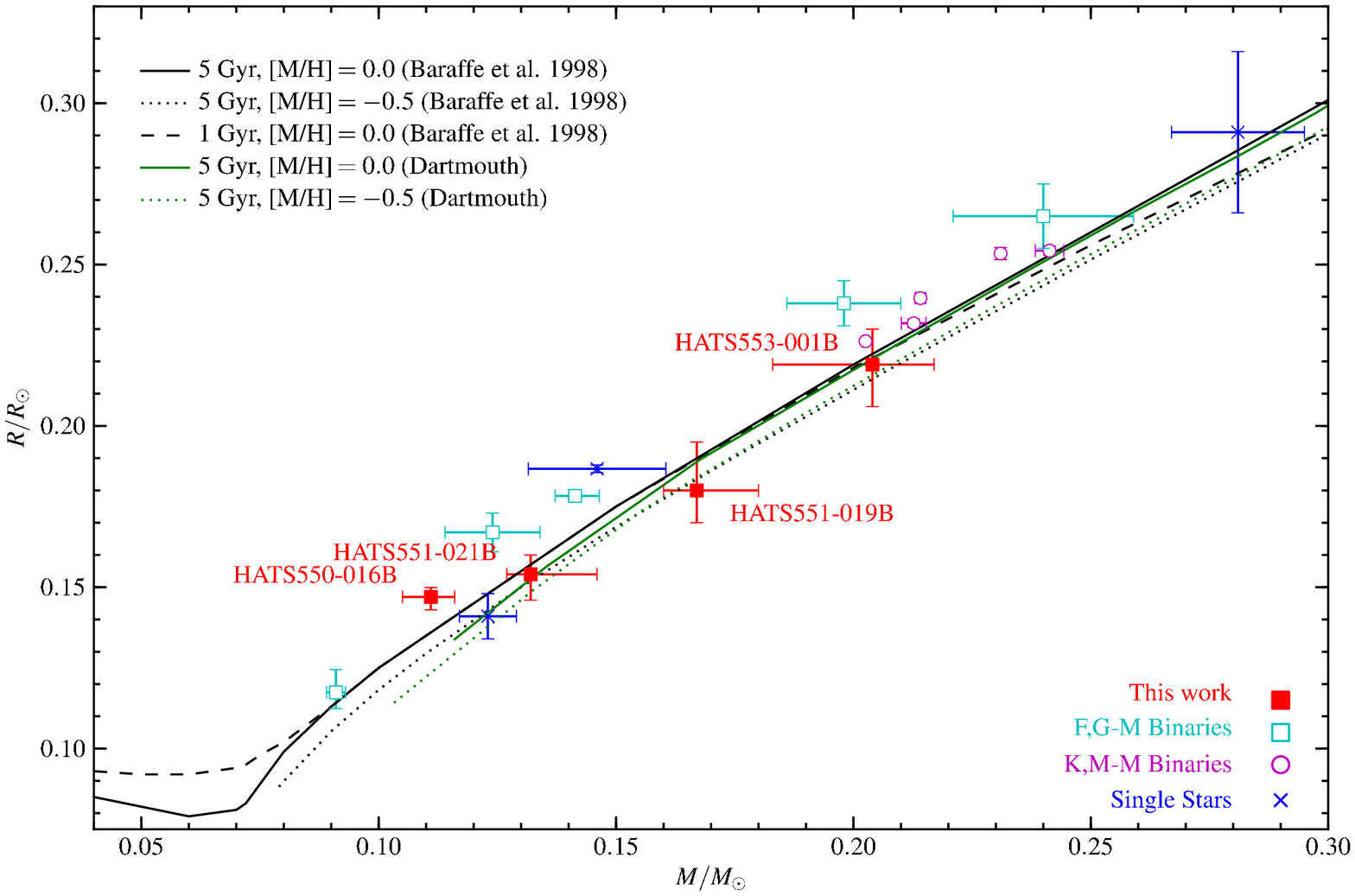}
  \caption{Mass--radius diagram of the VLMSs. New VLMSs presented in this paper are plotted in red and labelled, known F,G-M eclipsing binaries are plotted as cyan squares, M-M eclipsing binaries as magenta circles, and single stars measured by interferometry as blue crosses (Table~\ref{tab:prev_stars}). Isochrone lines from \citet{1998A&amp;A...337..403B} at 1.0 Gyr solar (black dashed), 5.0 Gyr solar (black solid), 5.0 Gyr $\text{[M/H]}=-0.5$ (black dotted), and from \citet{2008ApJS..178...89D} at 5.0 Gyr solar (green solid), 5.0 Gyr $\text{[M/H]}=-0.5$ (green dotted)  are marked. The corresponding 1 and 5 Gyr isochrones for brown dwarfs from \citet{2003A&amp;A...402..701B} are plotted in black, but not used in further analysis.}
  \label{fig:mass_radius}
\end{figure*}

The masses and radii of the VLMS companions presented in this paper, as well as that of known well-studied VLMSs (Table~\ref{tab:prev_stars}) are plotted in Figure~\ref{fig:mass_radius}. Of the VLMSs reported in this study, we find only HATS550-016B to be inflated compared to theoretical isochrones, HATS551-019B, HATS551-021B, and HATS553-001B agree with the isochrone mass-radius relations to within measurement uncertainties.

We tested for any general radius deviation between the observed VLMS population and the isochrones. The theoretical radii are taken from the 5 Gyr isochrones from \citet{1998A&amp;A...337..403B}, interpolated between $\text{[M/H]}=-0.5$ and 0.0, and linearly extrapolated beyond when necessary, to account for the metallicity dependency. For each object, we sample the isochrones via a Monte Carlo analysis, drawing mass and metallicity values from Gaussian distributions about the measured values and their associated uncertainties, and derive a predicted model radius and uncertainty. For this discussion on the radius deviation between model and measurements, we adopted the radius uncertainty as the quadrature addition of the uncertainties in the measurement and model sampling.

We also note that the difference between the \citet{1998A&amp;A...337..403B} and the Dartmouth isochrones \citep{2008ApJS..178...89D} is minor compared to the deviation from observations (see green isochrone lines in Figure~\ref{fig:mass_radius}), hence the following calculations were performed relative to the \citet{1998A&amp;A...337..403B} isochrones only. In addition, we also explored isochrones of younger ages and shorter mixing length using the \citet{1998A&amp;A...337..403B} isochrones, neither factors have obvious effects at this mass range.

The \chisq of the observed population is compared to (Model A) the isochrones without modification and (Model B) with isochrone radii inflated by 1.05. If the \chisq is calculated with the measurement uncertainties taken at face value, the Bayesean Information Criterion (BIC) between the two models is 50 ($\chi^2/\text{dof} = 7$ and 3, for Models A and B respectively), tentatively favouring an inflation of radius from the isochrones. The F-test for the variance ratio of the fit to the two models gives a p-value of 0.08, suggesting a very tentative preference towards Model B. Both BIC and the F-test account for the greater complexity of Model B over A. A number of studies \citep{2010ApJ...712.1003W,2010ApJ...718..502M,2012ApJ...757...42F} have commented that the presence of spots can impose radii uncertainties on the order of 2\% in these measurements. After imposing a minimum radius uncertainty of 2\% on the same population, we find $\text{BIC}=21$ ($\chi^2/\text{dof}=3$ and 2, for Models A and B respectively) and F-test p-value of 0.11. The F-test result, after inflating the uncertainties, suggests no real preference between Models A and B. In addition, 5\% systematic deviation between measurements and model is significantly smaller than the $\sim10$\% stated by earlier studies \citep[e.g.][]{2006Ap&amp;SS.304...89R}, and agrees with more recent studies of higher mass double-lined M-dwarf binaries, using newer isochrone sets \citep[e.g.][]{2011ApJ...728...48K,2012ApJ...757...42F,2013ApJ...776...87S}.

\begin{figure}
  \centering
  \includegraphics[width=9cm]{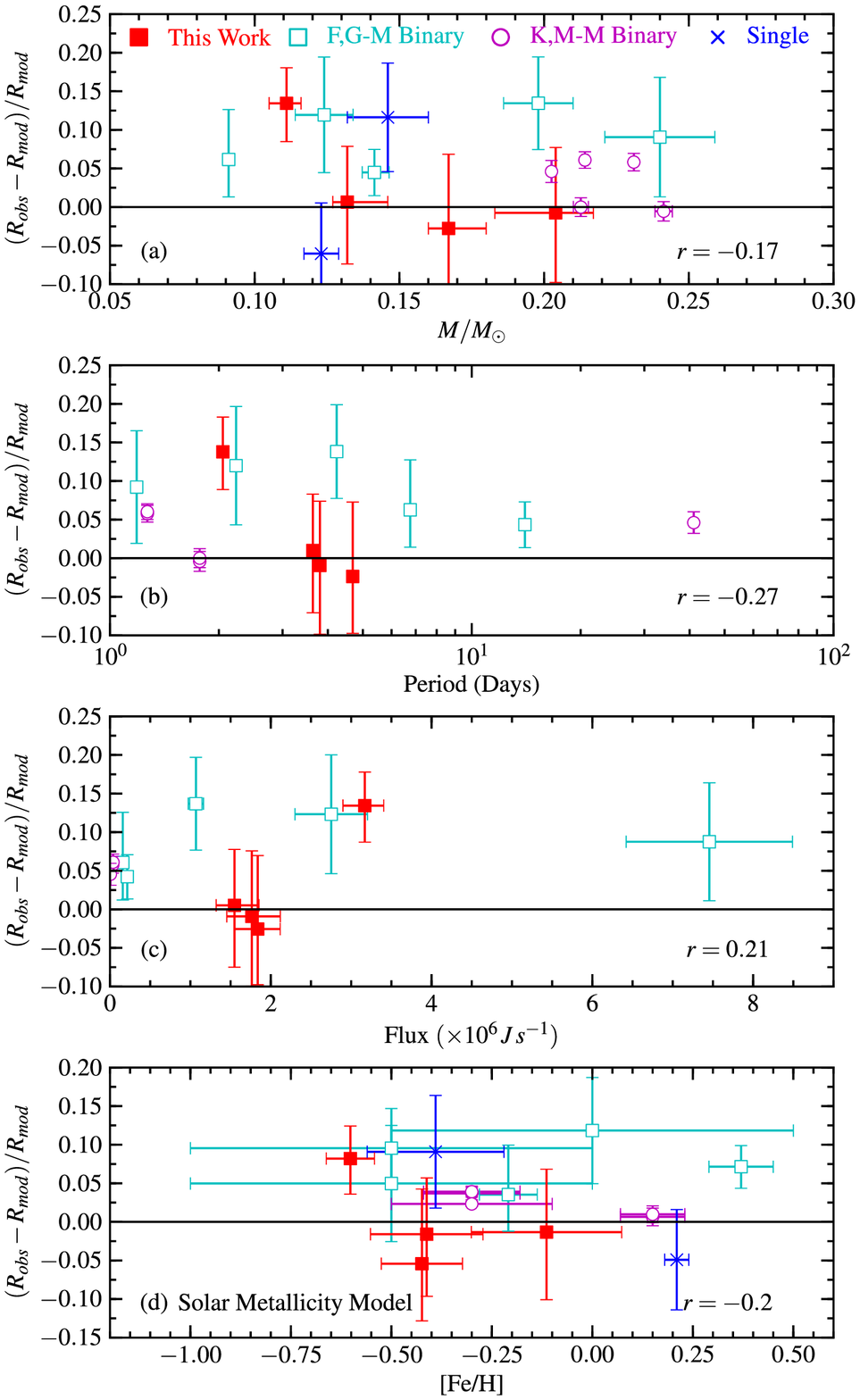}
  \caption{The $\text{observed}-\text{theoretical}$ radius difference of the VLMSs are plotted against their (a) mass, (b) system period, and (c) incident flux from the primary star. The theoretical radius is derived from the \citet{1998A&amp;A...337..403B} isochrones, interpolated to the [Fe/H] of each system. Objects without relevant orbital period and host star flux are omitted where necessary. Panel (d) shows the radius deviation to a solar metallicity isochrone as a function of the measured metallicity. The Pearson correlation coefficient $r$ is calculate for each panel. We find no significant correlation between the radius discrepancy and any of these parameters.}
  \label{fig:MR-trends}
\end{figure}

The RMS scatter of $\text{observed}-\text{theoretical}$ stellar radius difference (3\%) is slightly larger than the mean observational uncertainties (5\%). Whilst this is likely due to the under-estimated observational uncertainties, it may also point towards secondary factors, beyond mass and metallicity, that affect the radii of VLMSs. Figure~\ref{fig:MR-trends} plots the radius discrepancy against mass, orbital period, and incident flux from the primary star. For each factor, the Pearson correlation coefficient $r$ is calculated, weighted by the measurement uncertainties, with the radius uncertainties of M-M binaries increased to 2\%. We find no significant correlation with any of these parameters. 

In particular, activity induced inflation of the stellar radius should be correlated with shorter period if faster rotation gives rise to more powerful internal dynamos \citep{2007ApJ...660..732L}. This is not observed in the low mass population. However, it is not obvious that we expect such period-activity-radius dependencies, since the dynamos in fully convective stars may operate differently to solar type stars \citep{2006A&amp;A...446.1027C}. In addition, the paucity of well studied VLMS binaries in $>$10 day periods prevent us from drawing strong conclusions regarding the period--radius dependency. We also note the minimal dependency between incident flux and radius deviation. The influence of irradiation and disequilibrium chemistry on exoplanets has been contemplated \citep[e.g.][]{2010ApJ...720.1569K}, similar mechanisms that alter the top-level opacity of VLMSs may be possible, but clearly are not significant. In addition, whilst external influences are often considered when discussing the radius discrepancy of binary systems, we note that field M-dwarfs measured via interferometry have shown an equivalent radius discrepancy to the models \citep[e.g.][]{2006ApJ...644..475B,2012ApJ...757..112B,2013ApJ...776...87S}, and that there is no necessary expectation for radius, period, and incident flux to be correlated.

Metallicity has previously been suggested as a cause in the radius discrepancy \citep[e.g.][]{2006ApJ...644..475B,2011ApJ...736...47B}. Figure~\ref{fig:MR-trends} also plots the radius deviation to the solar metallicity isochrone. We find no correlation between the measured metallicity and the radius deviation. The lack of dependancy on metallicity for VLMSs agrees with the analysis by \citet{2013ApJ...776...87S}, who suggests that metallicity is only weakly correlated with radius, but more strongly affects luminosity and effective temperature. 

\input{literature_values.tex}

\section*{Acknowledgments}

Development of the HATSouth project was funded by NSF MRI grant
NSF/AST-0723074, operations are supported by NASA grant NNX09AB29G, and
follow-up observations receive partial support from grant
NSF/AST-1108686.
Work at the Australian National University is supported by ARC Laureate
Fellowship Grant FL0992131.
Followup observations with the ESO~2.2\,m/FEROS instrument were
performed under MPI guaranteed time (P087.A-9014(A), P088.A-9008(A),
P089.A-9008(A)).
AJ acknowledges support from FONDECYT project 1130857, BASAL CATA
PFB-06, and the Millenium Science Initiative, Chilean Ministry of
Economy (Nuclei: P10-022-F, P07-021-F).
RB and NE are supported by CONICYT-PCHA/Doctorado Nacional and MR
is supported by FONDECYT postdoctoral fellowship 3120097.
This work is based on observations made with ESO Telescopes at the La
Silla Observatory under programme IDs P087.A-9014(A), P088.A-9008(A),
P089.A-9008(A), P087.C-0508(A), 089.A-9006(A), and
This paper also uses observations obtained with facilities of the Las
Cumbres Observatory Global Telescope.
We acknowledge the use of the AAVSO Photometric All-Sky Survey (APASS),
funded by the Robert Martin Ayers Sciences Fund, and the SIMBAD
database, operated at CDS, Strasbourg, France.
Operations at the MPG/ESO 2.2\,m Telescope are jointly performed by the
Max Planck Gesellschaft and the European Southern Observatory.  The
imaging system GROND has been built by the high-energy group of MPE in
collaboration with the LSW Tautenburg and ESO \citep{2008PASP..120..405G}.  We thank Timo Anguita
and R\'egis Lachaume for their technical assistance during the
observations at the MPG/ESO 2.2\,m Telescope.
GZ thanks helpful discussions and draft proof readings by C.X. Huang.
\bibliographystyle{mn2e}
\bibliography{mybibfile}

\appendix
\FloatBarrier
\section{Light curve and radial velocity data}
\label{sec:appendix-data}

Tables ~\ref{tab:lctab-550-016}--\ref{tab:lctab-553-001} present the discovery and follow-up light curve data for the objects presented in this study.  Tables ~\ref{tab:rvtab-550-016}--\ref{tab:rvtab-553-001} present the associated radial velocity data.

\begin{table}
\begin{center}
\caption{
    Differential Photometry for HATS550-016 \label{tab:lctab-550-016}
}
{\footnotesize\begin{tabular}{lllrr}
\hline
    \multicolumn{1}{c}{BJD-2400000}          &
    \multicolumn{1}{c}{Flux} &
    \multicolumn{1}{c}{$\Delta$Flux} &
    \multicolumn{1}{c}{Instrument} &
    \multicolumn{1}{c}{Filter} \\
\hline
55103.463354 & 0.98794 & 0.01025 & HS & r' \\
55103.479516 & 1.00249 & 0.01028 & HS & r' \\
55103.485975 & 1.02298 & 0.00973 & HS & r' \\
55104.721430 & 0.99144 & 0.00734 & HS & r' \\
55130.407508 & 0.99268 & 0.00848 & HS & r' \\
... & ... & ... & ... & ...\\
\hline
\end{tabular}}
\end{center}
\begin{flushleft} 
This table is available in a machine-readable form in the online journal. A portion is shown here for guidance regarding its form and content.
\end{flushleft}
\end{table}

\begin{table}
\begin{center}
\caption{
    Differential Photometry for HATS551-019 \label{tab:lctab-551-019}
}
{\footnotesize\begin{tabular}{lllrr}
\hline
    \multicolumn{1}{c}{BJD-2400000}          &
    \multicolumn{1}{c}{Flux} &
    \multicolumn{1}{c}{$\Delta$Flux} &
    \multicolumn{1}{c}{Instrument} &
    \multicolumn{1}{c}{Filter} \\
\hline
55083.758920 & 1.00276 & 0.00311 & HS & r' \\
55083.762160 & 1.00041 & 0.00305 & HS & r' \\
55083.765520 & 1.00270 & 0.00303 & HS & r' \\
55083.768730 & 0.99539 & 0.00298 & HS & r' \\
55083.772100 & 1.00267 & 0.00296 & HS & r' \\
... & ... & ... & ... & ...\\
\hline
\end{tabular}}
\end{center}
\begin{flushleft} 
This table is available in a machine-readable form in the online journal. A portion is shown here for guidance regarding its form and content.
\end{flushleft}
\end{table}

\begin{table}
\begin{center}
\caption{
    Differential Photometry for HATS551-021 \label{tab:lctab-551-021}
}
{\footnotesize\begin{tabular}{lllrr}
\hline
    \multicolumn{1}{c}{BJD-2400000}          &
    \multicolumn{1}{c}{Flux} &
    \multicolumn{1}{c}{$\Delta$Flux} &
    \multicolumn{1}{c}{Instrument} &
    \multicolumn{1}{c}{Filter} \\
\hline
55083.758920 & 0.97941 & 0.00566 & HS & r' \\
55083.762160 & 0.99282 & 0.00577 & HS & r' \\
55083.765520 & 0.97866 & 0.00572 & HS & r' \\
55083.768730 & 0.98086 & 0.00569 & HS & r' \\
55083.772100 & 0.98239 & 0.00583 & HS & r' \\
... & ... & ... & ... & ...\\
\hline
\end{tabular}}
\end{center}
\begin{flushleft} 
This table is available in a machine-readable form in the online journal. A portion is shown here for guidance regarding its form and content.
\end{flushleft}
\end{table}

\begin{table}
\begin{center}
\caption{
    Differential Photometry for HATS553-001 \label{tab:lctab-553-001}
}
{\footnotesize\begin{tabular}{lllrr}
\hline
    \multicolumn{1}{c}{BJD-2400000}          &
    \multicolumn{1}{c}{Flux} &
    \multicolumn{1}{c}{$\Delta$Flux} &
    \multicolumn{1}{c}{Instrument} &
    \multicolumn{1}{c}{Filter} \\
\hline
55091.519969 & 0.98637 & 0.00568 & HS & r' \\
55091.523285 & 0.99027 & 0.00533 & HS & r' \\
55091.526738 & 0.99589 & 0.00528 & HS & r' \\
55091.530049 & 0.99446 & 0.00522 & HS & r' \\
55091.533508 & 0.98695 & 0.00512 & HS & r' \\
... & ... & ... & ... & ...\\
\hline
\end{tabular}}
\end{center}
\begin{flushleft} 
This table is available in a machine-readable form in the online journal. A portion is shown here for guidance regarding its form and content.
\end{flushleft}
\end{table}

\begin{table}
\begin{center}
\caption{
   Radial velocities for HATS550-016 \label{tab:rvtab-550-016}
}
{\footnotesize\begin{tabular}{lllr}
\hline
    \multicolumn{1}{c}{BJD-2400000}          &
    \multicolumn{1}{c}{RV (km\,s$^{-1}$)} &
    \multicolumn{1}{c}{$\Delta$RV (km\,s$^{-1}$)} &
    \multicolumn{1}{c}{Instrument} \\
\hline
56164.877923 & 21.49 & 0.08 & CORALIE \\
56237.817487 & 5.81 & 0.08 & CORALIE \\
56238.819306 & 16.33 & 0.08 & CORALIE \\
56239.850619 & 6.69 & 0.08 & CORALIE \\
56241.679708 & -3.25 & 0.08 & CORALIE \\
... & ... & ... & ...\\
\hline
\end{tabular}}
\end{center}
\begin{flushleft} 
This table is available in a machine-readable form in the online journal. A portion is shown here for guidance regarding its form and content.
\end{flushleft}
\end{table}

\begin{table}
\begin{center}
\caption{
   Radial velocities for HATS551-019 \label{tab:rvtab-551-019}
}
{\footnotesize\begin{tabular}{lllr}
\hline
    \multicolumn{1}{c}{BJD-2400000}          &
    \multicolumn{1}{c}{RV (km\,s$^{-1}$)} &
    \multicolumn{1}{c}{$\Delta$RV (km\,s$^{-1}$)} &
    \multicolumn{1}{c}{Instrument} \\
\hline
55497.745634 & 10.14 & 0.04 & CORALIE \\
55608.701137 & 33.44 & 0.04 & CORALIE \\
55610.699243 & -0.51 & 0.05 & CORALIE \\
56374.907520 & -7.46 & 0.61 & ANU2.3M/ECHELLE \\
56375.887520 & 4.32 & 0.56 & ANU2.3M/ECHELLE \\
... & ... & ... & ...\\
\hline
\end{tabular}}
\end{center}
\begin{flushleft} 
This table is available in a machine-readable form in the online journal. A portion is shown here for guidance regarding its form and content.
\end{flushleft}
\end{table}

\begin{table}
\begin{center}
\caption{
   Radial velocities for HATS551-021 \label{tab:rvtab-551-021}
}
{\footnotesize\begin{tabular}{lllr}
\hline
    \multicolumn{1}{c}{BJD-2400000}          &
    \multicolumn{1}{c}{RV (km\,s$^{-1}$)} &
    \multicolumn{1}{c}{$\Delta$RV (km\,s$^{-1}$)} &
    \multicolumn{1}{c}{Instrument} \\
\hline
55497.766487 & 28.16 & 0.06 & CORALIE \\
55608.553710 & -1.30 & 0.07 & CORALIE \\
55610.627251 & 26.18 & 0.06 & CORALIE \\
55813.836900 & 28.69 & 0.28 & FEROS \\
55814.854700 & 8.32 & 0.25 & FEROS \\
... & ... & ... & ...\\
\hline
\end{tabular}}
\end{center}
\begin{flushleft} 
This table is available in a machine-readable form in the online journal. A portion is shown here for guidance regarding its form and content.
\end{flushleft}
\end{table}

\begin{table}
\begin{center}
\caption{
   Radial velocities for HATS553-001 \label{tab:rvtab-553-001}
}
{\footnotesize\begin{tabular}{lllr}
\hline
    \multicolumn{1}{c}{BJD-2400000}          &
    \multicolumn{1}{c}{RV (km\,s$^{-1}$)} &
    \multicolumn{1}{c}{$\Delta$RV (km\,s$^{-1}$)} &
    \multicolumn{1}{c}{Instrument} \\
\hline
56374.955480 & 26.75 & 0.80 & ANU2.3M/ECHELLE \\
56375.925980 & -3.30 & 0.71 & ANU2.3M/ECHELLE \\
56377.923100 & 25.50 & 0.74 & ANU2.3M/ECHELLE \\
56378.918650 & 23.46 & 0.73 & ANU2.3M/ECHELLE \\
56380.938390 & -2.56 & 1.24 & ANU2.3M/ECHELLE \\
... & ... & ... & ...\\
\hline
\end{tabular}}
\end{center}
\begin{flushleft} 
This table is available in a machine-readable form in the online journal. A portion is shown here for guidance regarding its form and content.
\end{flushleft}
\end{table}

\FloatBarrier

\label{lastpage}

\end{document}

%% file: params_table.tex
\begin{table*}
\begin{center}
\caption{Properties of the HATSouth Transiting VLMS systems
\label{tab:stellar-params}}
{\footnotesize\begin{tabular}{lrrrr}
\hline
\multicolumn{1}{c}{Parameter} & 
\multicolumn{1}{c}{HATS550-016} &
\multicolumn{1}{c}{HATS551-019} &
\multicolumn{1}{c}{HATS551-021} &
\multicolumn{1}{c}{HATS553-001} \\ 
\hline
\multicolumn{2}{l}{\textbf{Fitted parameters}}\\ 
P (days) & $2.051811_{-0.000002}^{+0.000002}$ & $4.68681_{-0.00001}^{+0.00002}$ & $3.63637_{-0.00005}^{+0.00005}$ & $3.80405_{-0.00001}^{+0.00001}$ \\ 
$T_0$ (HJD) & $2455104.286_{-0.001}^{+0.001}$ & $2455474.179_{-0.002}^{+0.001}$ & $2455087.426_{-0.003}^{+0.002}$ & $2455093.563_{-0.001}^{+0.002}$ \\ 
$R_\text{sum}$ & $0.196_{-0.004}^{+0.003}$ & $0.149_{-0.008}^{+0.006}$ & $0.131_{-0.005}^{+0.004}$ & $0.158_{-0.007}^{+0.005}$ \\ 
$R_2/R_1$ & $0.1205_{-0.0003}^{+0.0003}$ & $0.107_{-0.002}^{+0.002}$ & $0.124_{-0.002}^{+0.003}$ & $0.136_{-0.004}^{+0.003}$ \\ 
i \, $(^\circ)$ & $90_{-1}^{+1}$ & $85_{-1}^{+1}$ & $90_{-1}^{+1}$ & $83.4_{-0.3}^{+0.4}$ \\ 
$e\cos \omega$ & $-0.001_{-0.002}^{+0.003}$ & $-0.003_{-0.005}^{+0.002}$ & $0.003_{-0.002}^{+0.002}$ & $0.000_{-0.002}^{+0.001}$ \\ 
$e\sin \omega$ & $0.08_{-0.02}^{+0.02}$ & $0.04_{-0.02}^{+0.02}$ & $0.06_{-0.02}^{+0.02}$ & $0.03_{-0.02}^{+0.02}$ \\ 
$K\,(\text{kms}^{-1})$ & $17.7_{-0.5}^{+0.4}$ & $18.4_{-0.7}^{+0.6}$ & $16.3_{-0.2}^{+0.2}$ & $20.9_{-0.9}^{+0.8}$ \\ 
\\ 
 \multicolumn{2}{l}{\textbf{Derived parameters}}\\ 
$\log g$ & $4.25_{-0.02}^{+0.02}$ & $4.01_{-0.05}^{+0.05}$ & $4.30_{-0.04}^{+0.04}$ & $4.13_{-0.05}^{+0.05}$ \\ 
Age (Gyr) & $5_{-1}^{+3}$ & $6_{-2}^{+2}$ & $4_{-4}^{+3}$ & $3_{-2}^{+2}$ \\ 
$M_1 \, (M_\odot)$ & $0.97_{-0.06}^{+0.05}$ & $1.10_{-0.09}^{+0.05}$ & $1.1_{-0.1}^{+0.1}$ & $1.2_{-0.1}^{+0.1}$ \\ 
$R_1 \, (R_\odot)$ & $1.22_{-0.03}^{+0.02}$ & $1.70_{-0.09}^{+0.09}$ & $1.20_{-0.01}^{+0.08}$ & $1.58_{-0.03}^{+0.08}$ \\ 
$M_2 \, (M_\odot)$ & $0.110_{-0.006}^{+0.005}$ & $0.17_{-0.01}^{+0.01}$ & $0.132_{-0.005}^{+0.014}$ & $0.20_{-0.02}^{+0.01}$ \\ 
$R_2 \, (R_\odot)$ & $0.147_{-0.004}^{+0.003}$ & $0.18_{-0.01}^{+0.01}$ & $0.154_{-0.008}^{+0.006}$ & $0.22_{-0.01}^{+0.01}$ \\ 
\hline
\end{tabular}}
\end{center}
\end{table*}

%% file: system_plots.tex
\begin{figure}
  \centering
  \includegraphics[width=9cm]{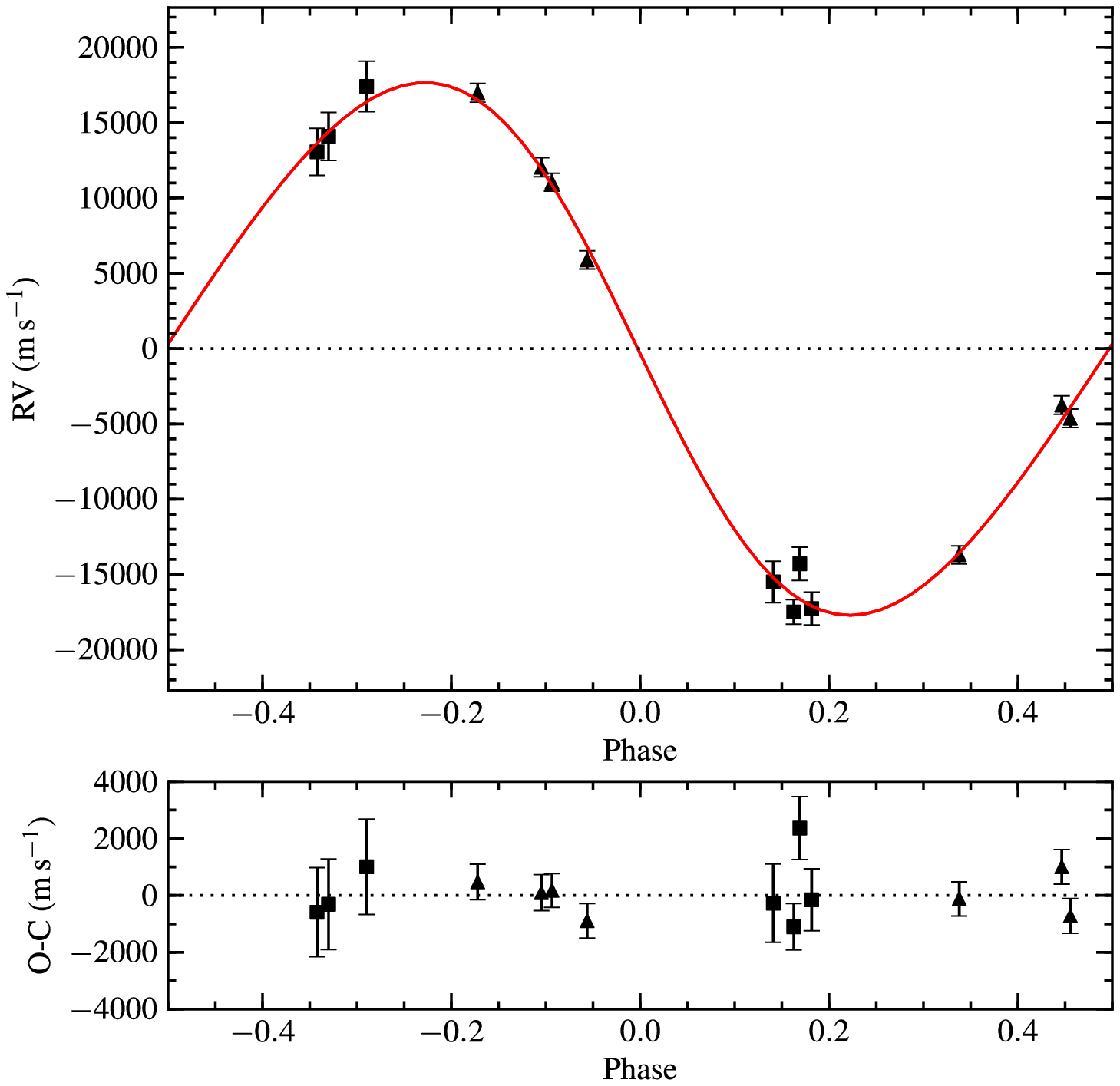}\\
  \includegraphics[width=9cm]{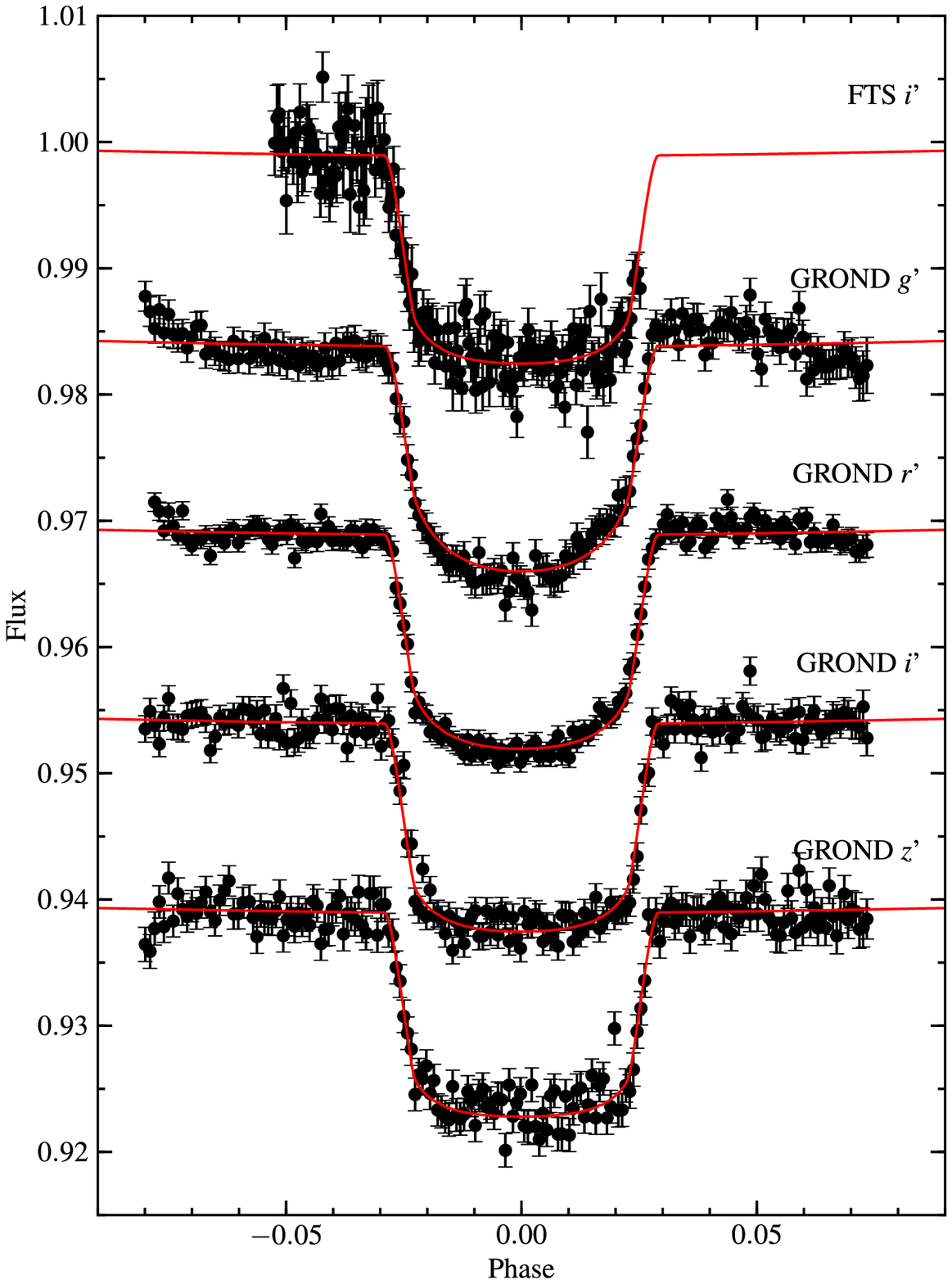}
  \caption{Top: HATS550-016 radial velocities and the Keplerian orbit fit. ANU 2.3\,m Echelle data are plotted as squares, Coralie data as triangles. Bottom: Follow-up transit light curves and model fit.}
  \label{fig:fit_550-016}
\end{figure}

\begin{figure}
  \centering
  \includegraphics[width=9cm]{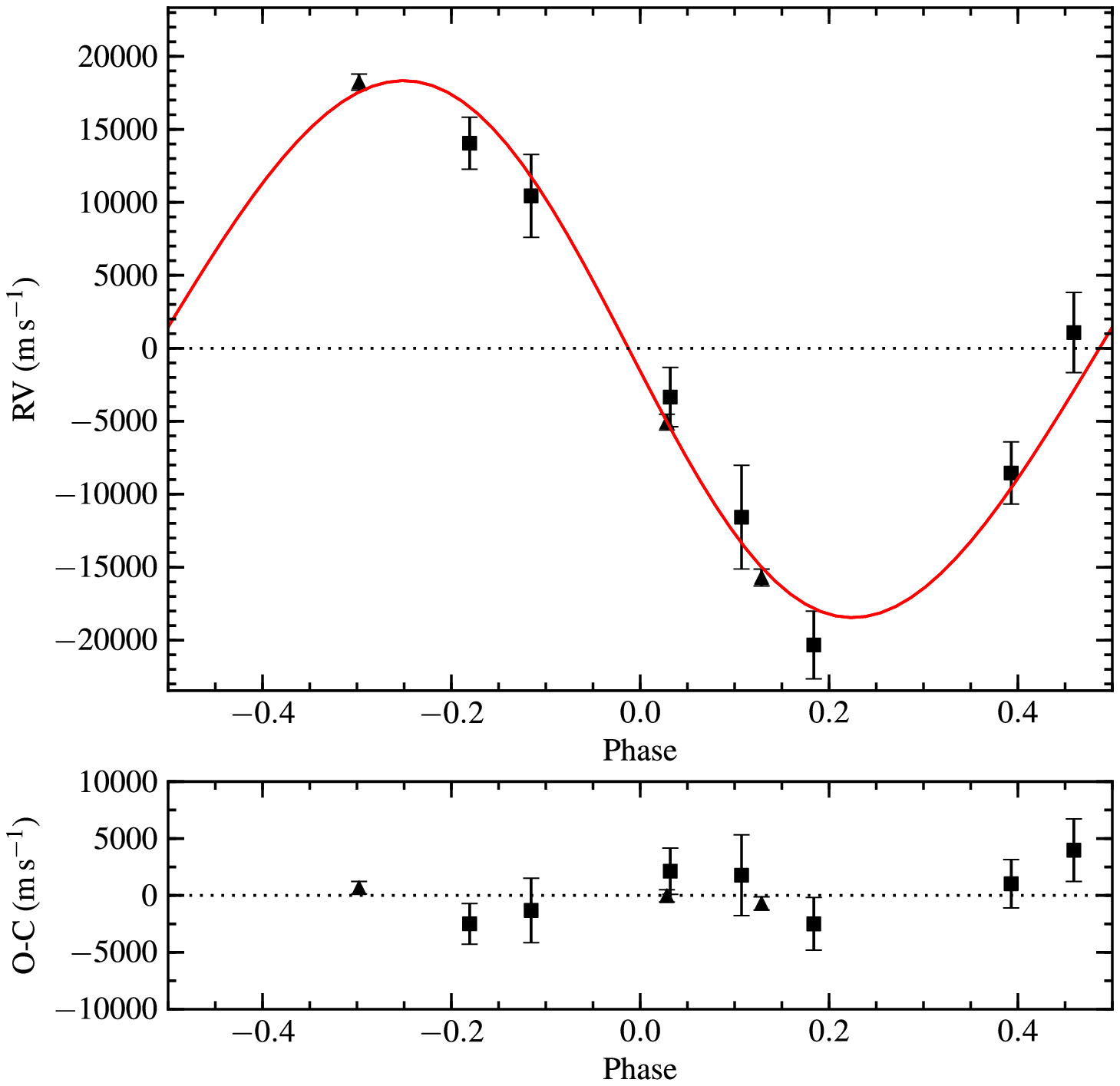}\\
  \includegraphics[width=9cm]{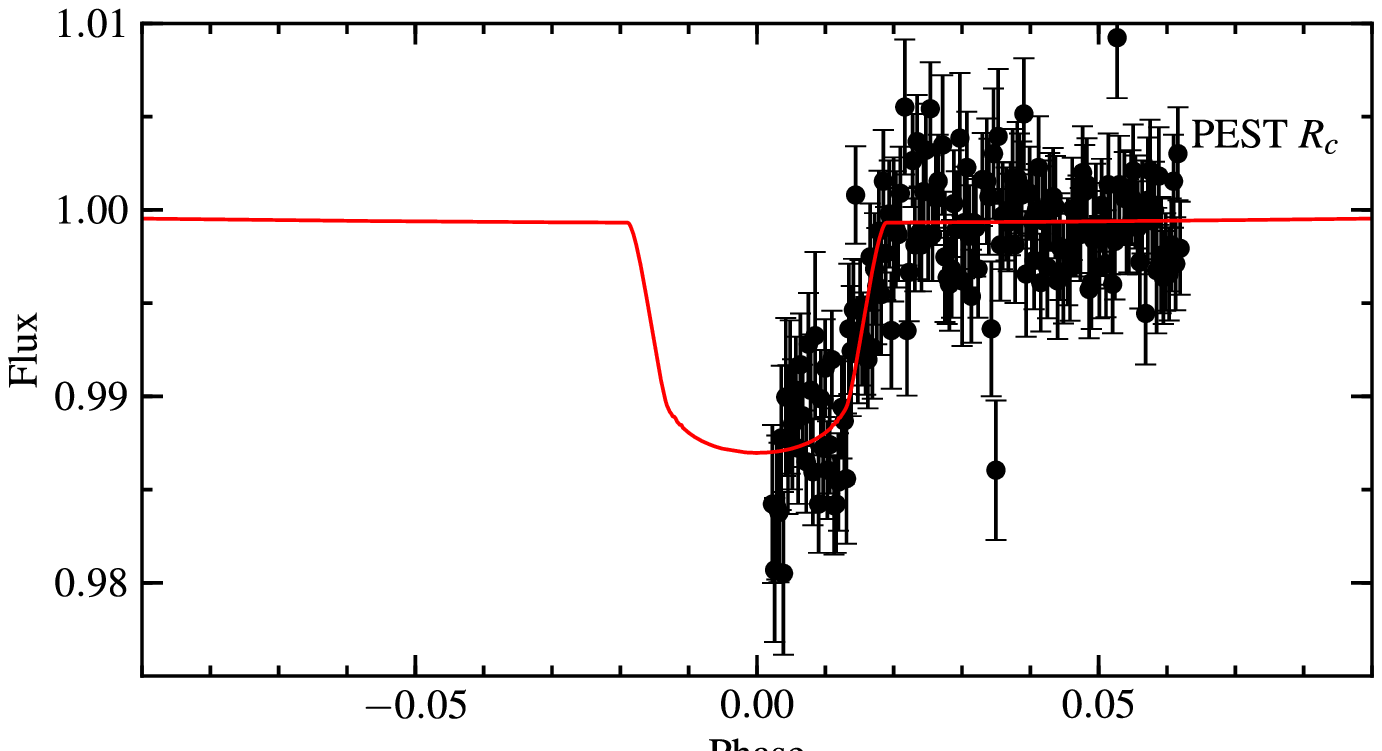}
  \caption{Top: HATS551-019 radial velocities and the Keplerian orbit fit. ANU 2.3\,m Echelle data are plotted as squares, Coralie data as triangles. Bottom: Follow-up transit light curve and model fit. }
  \label{fig:fit_551-019}
\end{figure}

\begin{figure}
  \centering
  \includegraphics[width=9cm]{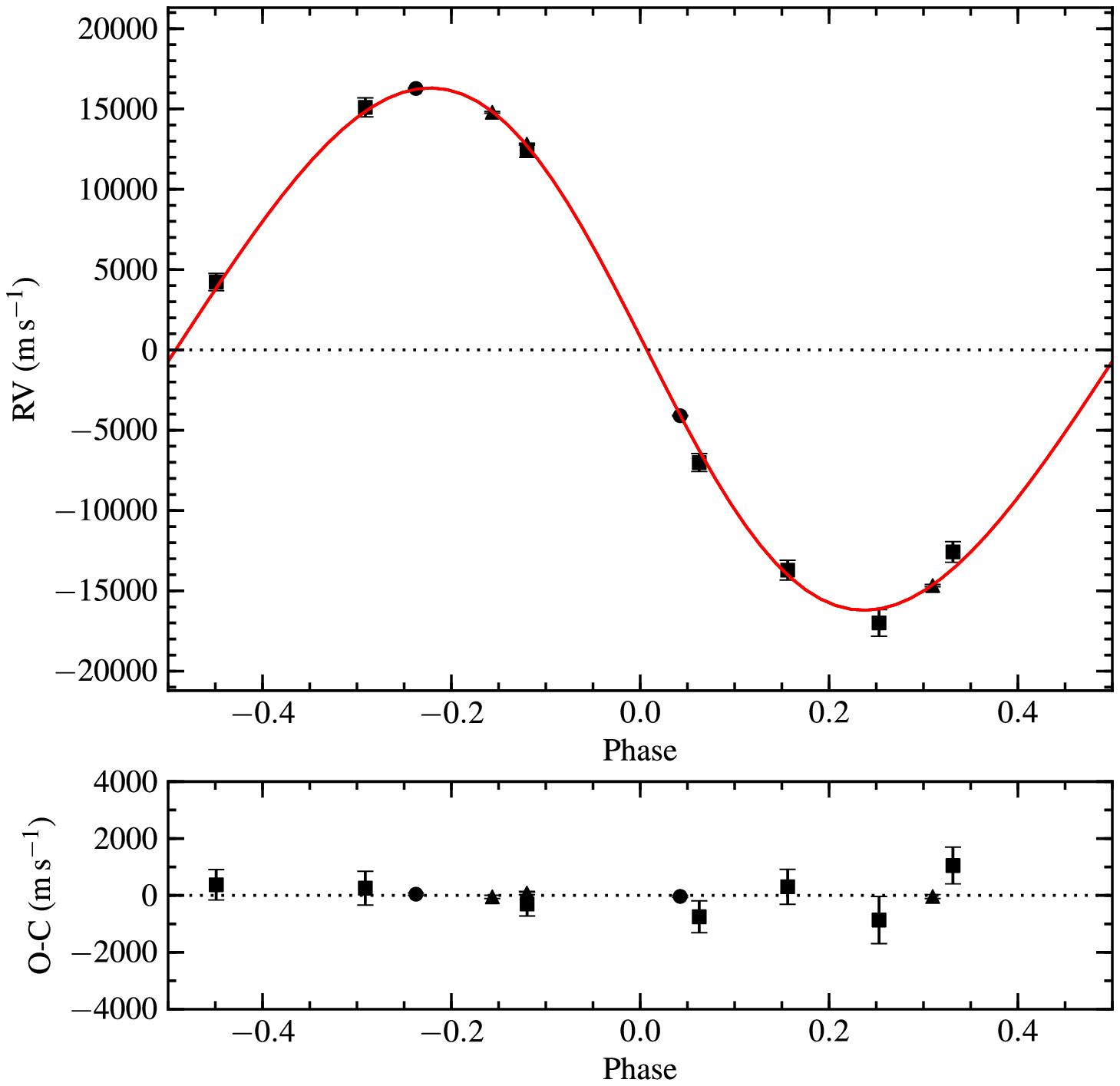}\\
  \caption{HATS551-021 radial velocities and the Keplerian orbit fit. ANU 2.3\,m Echelle data are plotted as squares, Coralie data as triangles, FEROS data as circles.}
  \label{fig:fit_551-021}
\end{figure}

\begin{figure}
  \centering
  \includegraphics[width=9cm]{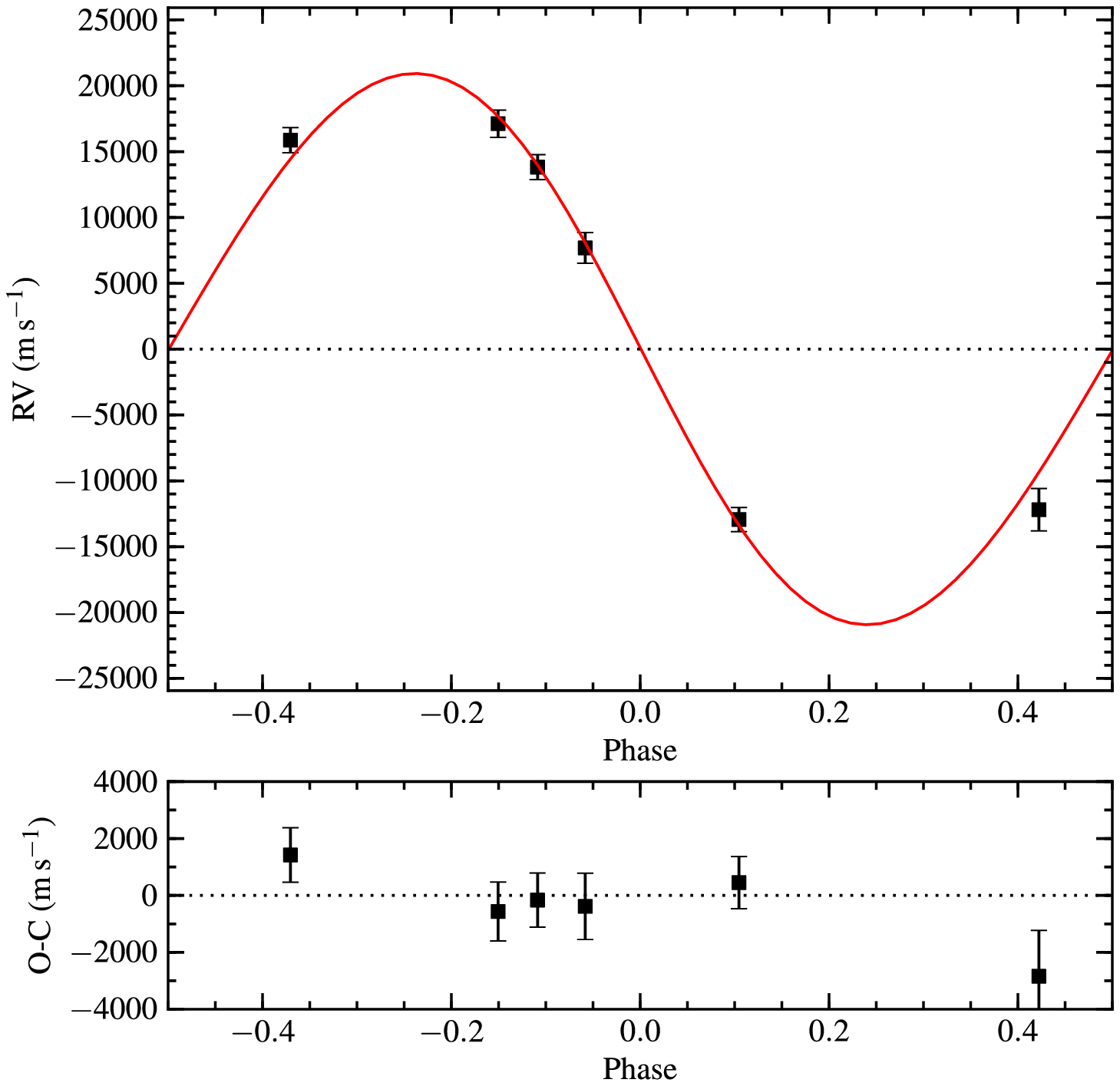}\\
  \includegraphics[width=9cm]{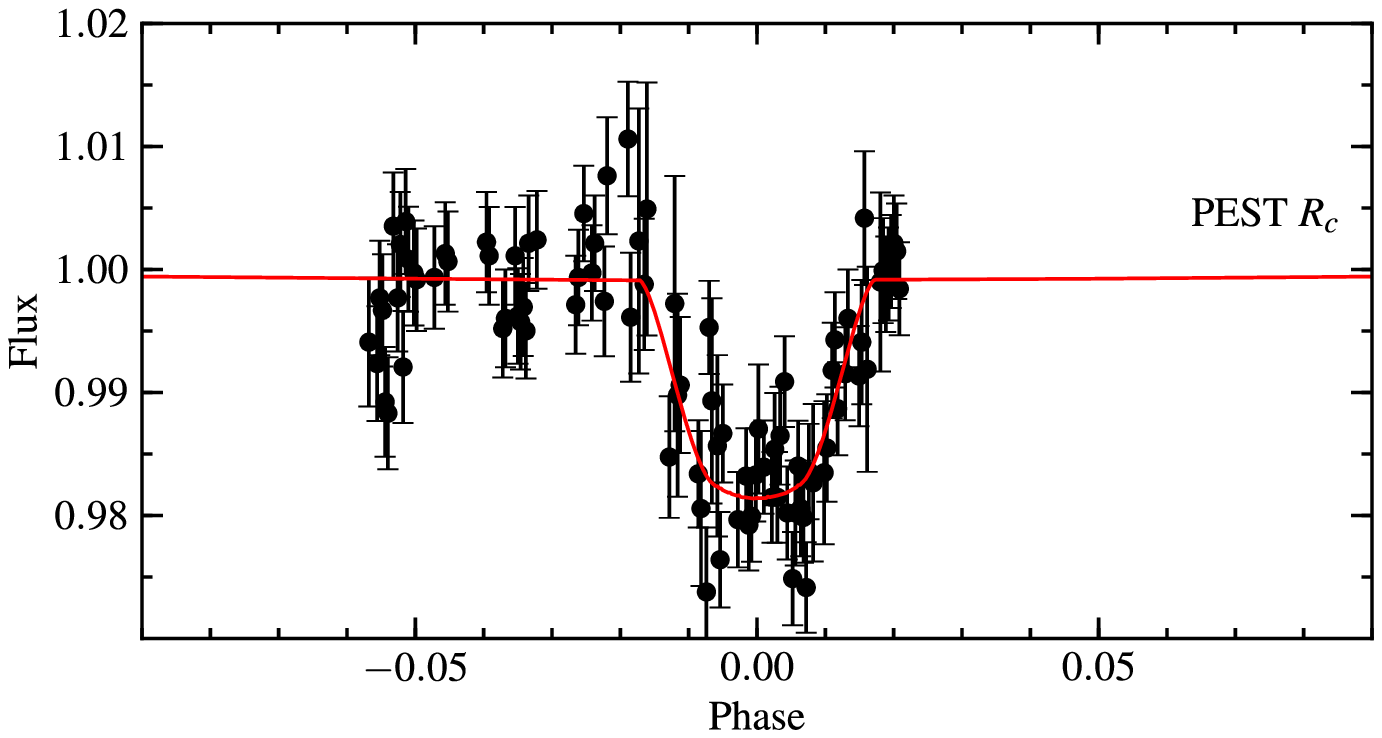}
  \caption{Top: HATS553-001 radial velocities and the Keplerian orbit fit. ANU 2.3\,m Echelle data are plotted as squares. Bottom: Follow-up transit light curve and model fit. }
  \label{fig:fit_553-001}
\end{figure}

%% file: literature_values.tex
\begin{landscape}
\begin{table}

\begin{center}
\caption{
    Properties of known VLMSs$^{a}$ \label{tab:prev_stars}
}
{\footnotesize\begin{tabular}{lrrrrrrrl}
\hline
    \multicolumn{1}{c}{Object}          &
    \multicolumn{1}{c}{Mass $(M_\odot)$} &
    \multicolumn{1}{c}{Radius $(R_\odot)$} &
    \multicolumn{1}{c}{Method} &
    \multicolumn{1}{c}{[Fe/H]} &
    \multicolumn{1}{c}{Period (days)} &
    \multicolumn{1}{c}{ $T_\text{eff}\, (\text{K})$} &
    \multicolumn{1}{c}{Companion} &
    \multicolumn{1}{c}{Reference} \\
    & & & & & & & \multicolumn{1}{c}{$T_\text{eff}\, (\text{K})$} & \\
\hline
\multicolumn{5}{l}{\bf{F,G-M Binaries}$^b$}\\
HAT-TR-205-013B & $0.124\pm0.01$ & $0.167\pm0.006$ & SB1, synchronisation$^d$ & $0.0\pm0.5$ & 2.23 & & $6295\pm200$ & \citet{2007ApJ...663..573B}\\
J1219-39B & $0.091\pm0.002$ & $0.1174_{-0.0050}^{+0.0071}$ & SB1, isochrone$^c$ & $-0.209 \pm 0.072$ & 6.76 & & $5400\pm90$ & \citet{2013Aamp;A...549A..18T}\\
KIC 1571511B & $0.14136_{-0.0042}^{+0.0051}$ & $0.17831_{-0.0016}^{+0.0013}$ & SB1, isochrone & $0.37\pm0.08$ & 14.02 & $4090 \pm 60$ & $6195\pm50$ & \citet{2012MNRAS.423L...1O}\\
T-Lyr0-08070B & $0.240\pm0.019$ & $0.265\pm0.010$ & SB1, synchronisation & $-0.5^e$ & 1.18 & & $6250\pm140$ & \citet{2009ApJ...701..764F}\\
T-Lyr1-01662B & $0.198\pm0.012$ & $0.238\pm0.007$ & SB1, synchronisation & $-0.5^e$ & 4.23 & & $6200\pm30$ & \citet{2009ApJ...701..764F}\\
\\
\multicolumn{5}{l}{\bf{K,M-M Binaries}}\\
CM Dra A & $0.2130\pm0.0009$ & $0.2534\pm0.0019$ & SB2$^f$ & $-0.3 \pm 0.12$ & 1.27 & $3130\pm70$ & $3120\pm70$ & \citet{2009ApJ...691.1400M}\\ &&&&&&&&\citet{2012ApJ...760L...9T}\\
CM Dra B & $0.2141\pm0.0010$ & $0.2396\pm0.0015$ & SB2 & $-0.3 \pm 0.12$ & 1.27 & $3120\pm70$ & $3130\pm70$ & \citet{2009ApJ...691.1400M}\\ &&&&&&&&\citet{2012ApJ...760L...9T}\\
Kepler-16B & $0.20255_{-0.00065}^{+0.00066}$ & $0.22623_{-0.00053}^{+0.00059}$ & SB1, photodynamical$^g$ & $-0.3\pm0.2$ & 41.08 & & $4450\pm150$ & \citet{2011Sci...333.1602D}\\
KOI-126B & $0.2413\pm0.003$ & $0.2543\pm0.0014$ & SB1, photodynamical & $0.15\pm0.08$ & 1.77 & & $5875\pm100$ & \citet{2011Sci...331..562C}\\
&&&&& (About KOI-126C) && (KOI-126A)&\\
&&&&& 33.92 &&&\\
&&&&& (About KOI-126A) &&&\\
KOI-126C & $0.2127\pm0.0026$ & $0.2318\pm0.0013$ & SB1, photodynamical & $0.15\pm0.08$ & 1.77 & & $5875\pm100$ & \citet{2011Sci...331..562C}\\
&&&&& (About KOI-126B) && (KOI-126A)&\\
&&&&& 33.92 &&&\\
&&&&& (About KOI-126A) &&&\\
\\
\multicolumn{5}{l}{\bf{Single Stars}}\\
GJ 191 & $0.281\pm0.014$ & $0.291\pm0.025$ & Interferometry & $-0.99\pm0.04$ & & $3570\pm160$ & & \citet{2003Aamp;A...397L...5S}\\ &&&&&&&&\citet{2005MNRAS.356..963W}\\
GJ 551 & $0.123 \pm 0.006$ & $0.141 \pm 0.007$ & Interferometry & $0.21\pm0.03^h$ & & $3098\pm56$ & & \citet{2003Aamp;A...397L...5S} \\ &&&&&&&& \citet{2005ApJS..159..141V}\\ &&&&&&&& \citet{2009Aamp;A...505..205D} \\
GJ 699 & $0.146\pm0.015$ & $0.1867\pm0.0012$ & Interferometry & $-0.39\pm0.17$ & & $3224\pm10$ & & \citet{2012ApJ...757..112B}\\ &&&&&&&&\citet{2012ApJ...748...93R}\\
\hline
\end{tabular}}
\end{center}
\begin{flushleft} 
$^{a}${With $0.08 <M < 0.3\,M_\odot$, mass and radius measured to better than 10\% precision, and valid [Fe/H] measurements}\\
$^{b}${Assume primary star [Fe/H]}\\
$^{c}${SB1, isochrone: Single lined stellar binary, parameters derived from isochrone fitting}\\
$^{d}${SB1, synchronisation: Single lined stellar binary, parameters derived from assuming spin-orbit synchronisation}\\
$^{e}${[Fe/H] adopted from Table 13 of \citet{2009ApJ...701..764F}, by finding the best matching results between the isochrone and synchronisation techniques. We assume an error of 0.5 dex (grid size) in our analysis of BIC and F-test.}\\
$^{f}${SB2: Double lined eclipsing binary, parameters determined dynamically.}\\
$^{g}${SB1, photodynmical: Global analysis of single-lined radial velocity data and light curve transit timing variations for multi-body systems.}\\
$^{h}${Adopting [Fe/H] of $\alpha$ Centauri A, see \citet{2009ApJ...699..933J}.}
\end{flushleft}
\end{table}
\end{landscape}